\documentclass[a4paper,12pt,dvipsnames,usernames]{article}
\pdfoutput=1

\usepackage{jheppub} 

\usepackage[utf8]{inputenc}
\usepackage{graphicx}
\usepackage{amsmath}
\usepackage{amssymb}
\usepackage{bm}
\usepackage{paralist,array}
\usepackage{units} 
\graphicspath{{fig/}{./Plots/}} 
\usepackage{slashed}
\usepackage{todonotes}
\usepackage[parfill]{parskip}
\usepackage{longtable}
\usepackage{subfig}

\title{Production of  Chern-Simons bosons in  decays of mesons}

\author[1]{Yuliia  Borysenkova,}
\author[1,2]{Pavlo  Kashko,}
\author[1]{Mariia  Tsarenkova,}
\author[3,4]{Kyrylo~Bondarenko,} 
\author[1]{Volodymyr~Gorkavenko}

\affiliation[1]{Department of Physics, Taras Shevchenko National University of Kyiv, 64 Volodymyrs’ka str. 01601, Kyiv, Ukraine}
\affiliation[2]{\'{E}cole Polytechnique F\'{e}d\'{e}rale de Lausanne (EPFL),
CH-1015 Lausanne, Switzerland}
\affiliation[3]{SISSA, International School for Advanced Studies, Via Bonomea 265, I-34136 Trieste, Italy}
\affiliation[4]{IFPU, Institute for Fundamental Physics of the Universe, via Beirut 2, 34151, Trieste, Italy}

\emailAdd{yuliya.borisenckova@gmail.com}
\emailAdd{kashko.pavlo@gmail.com}
\emailAdd{ters.mar@gmail.com}
\emailAdd{kyrylo.bondarenko@gmail.com}
\emailAdd{gorkavol@gmail.com}

\begin{document}

\abstract{We consider the effective interaction of quarks with a new GeV-scale vector particle  that couples to electroweak gauge bosons by the so-called effective Chern-Simons interaction. We call this particle the Chern-Simons (CS) boson. We construct effective Lagrangian of the CS boson interaction with quarks of two different flavors. This interaction is given by a divergent loop diagram, however, it turns out that the divergent part is equal to zero as a consequence of the CKM matrix unitarity in the SM. Therefore, we are able to predict effective interaction of the CS boson with quarks of different flavors without introducing new unknown parameters to the model, using only parameters of the initial effective Lagrangian. Our result shows that the effective interaction of the CS boson with down-type quarks is sufficiently stronger compared with up-type quarks. Based on our results, we give a prediction for the production of CS bosons in mesons decays. Branching fractions were obtained for the main reactions of the CS production in meson decays. 
The results obtained will be useful for searching for the long-lived GeV-scale CS boson in intensity frontier experiments.}

\maketitle

\newpage

\section{Introduction}\label{sec:intro}

Despite all the successes  of the Standard Model (SM), see e.g. \cite{Altarelli:2013tya}, there are some phenomena  that can not be solved within the SM. These include baryon asymmetry of the Universe (see e.g. \cite{Steigman:1976ev,Riotto:1999yt,Canetti:2012zc}),  dark matter (see e.g. \cite{Peebles:2013hla,Lukovic:2014vma,Bertone:2016nfn}), and neutrino oscillations (see e.g. \cite{Bilenky:1987ty,Strumia:2006db,deSalas:2017kay}). 
In addition to these well-established phenomena, it should be noted that there is a number of observed parameters that are difficult to explain. For example, the strong CP problem (as to why the degree of CP violation in the QCD is unobservably small, see e.g. \cite{Czarnecki:1997bu,Kim:2008hd}), the Higgs hierarchy problem (as to why quantum corrections to the Higgs mass cancel well, see e.g. \cite{Schmaltz:2005ky,Wells:2016luz}), stability of the SM vacuum (top quark Yukawa coupling and the Higgs mass are very close to its critical value, see e.g. \cite{Degrassi:2012ry,Bezrukov:2014ina}), the cosmological constant and dark energy (as to why the cosmological constant is so small, see e.g. \cite{Padmanabhan:2002ji}).
Therefore, one can conclude that the SM is an incomplete theory and it requires an extension.
Moreover, the existence of "hidden" sectors with particles of new physics seems plausible. 

Many of the aforementioned phenomena can be explained by an extension of the SM by Beyond Standard Model (BSM)  particles. The values of their masses can be in very different ranges. For example, small neutrino masses, dark matter, and baryon asymmetry of the Universe can be explained by new particles with masses from the sub-eV scale up to the GUT scale, see e.g. \cite{Strumia:2006db, Rubakov:2017xzr}. The fact that we do not observe BSM particles  in accelerator experiments has two possible explanations. Either these particles are too massive to be produced at modern accelerators like the LHC, or they feebly interact with SM particles.
If BSM particles are heavy enough, 
 then to search for them we need more powerful, more expensive  accelerators and energy frontier experiments, see e.g. \cite{Golling:2016gvc,FCC:2018byv}. But the case of  light but feebly interacting BSM particles  is very topical
for the experimental search for new physics right now. To search for them, we need intensity frontier experiments with high-intensity particle beams and large detectors, see e.g. \cite{Beacham:2019nyx,Lanfranchi:2020crw}.  
Several such intensity frontier experiments have been proposed in recent years:  MATHUSLA \cite{Curtin:2018mvb}, FACET \cite{Cerci:2021nlb}, FASER \cite{FASER:2018ceo,FASER:2018eoc}, SHiP \cite{Anelli:2015pba,Alekhin:2015byh}, NA62 \cite{Mermod:2017ceo,NA62:2017qcd,Drewes:2018gkc}, DUNE \cite{DUNE:2015lol,DUNE:2020fgq}, etc.

Searching for new physics, one should keep in mind that among the hidden particles there must be particles that solve the above-mentioned SM problems. However, the hidden sector may also contain particles that are not directly related to the solution of an SM problem. 
That is why it is important to look for manifestations of  new physics particles at all accessible energy scales. 
In this work we will be interested in GeV-scale feebly interacting new particles. 

The properties of the new light particles are not yet known. They can be  scalars (e.g. \cite{Patt:2006fw,Bezrukov:2009yw,Boiarska:2019jym}),  pseudoscalars (axion-like particles), see e.g. \cite{Peccei:1977hh, Weinberg:1977ma, Wilczek:1977pj,Choi:2020rgn},  vectors (e.g. \cite{Okun:1982xi,Holdom:1985ag,Langacker:2008yv}), or  fermions (e.g. \cite{Asaka:2005pn,Asaka:2005an,Bondarenko:2018ptm,Boyarsky:2018tvu}). Each of these options requires thorough study.
Discussion of their possible interactions with SM particles (portals) and the available experimental constraints are given e.g. in reviews
\cite{Alekhin:2015byh,Curtin:2018mvb}.

In this paper, we will be interested in consideration of the Chern-Simons portal, which, in our opinion, has received insufficient attention especially concerning experimental search for the Chern-Simons particles.  
In this portal a
new vector particle $X_\mu$ couples with the SM gauge fields by the so-called
effective Chern-Simons interaction in  the  form  of  4-dimension operators \cite{Antoniadis:2009ze}: 
\begin{equation}\label{Lcs}  
     \mathcal{L}_{CS}=c_z \epsilon^{\mu\nu\lambda\rho} X_\mu Z_\nu \partial_\lambda Z_\rho +c_\gamma \epsilon^{\mu\nu\lambda\rho} X_\mu Z_\nu \partial_\lambda A_\rho+\left\{ c_w \epsilon^{\mu\nu\lambda\rho} X_\mu W_\nu^- \partial_\lambda W_\rho^+ + h.c.\right\},
\end{equation}
where $A_\mu$ is the electromagnetic field; $W^\pm_\mu$ and $Z_\mu$ are  fields of the weak interaction; $\epsilon^{\mu\nu\lambda\rho}$ is the Levi-Civita symbol ($\epsilon^{0123}=+1$) and $c_z$, $c_\gamma$, $c_w$ are some dimensionless independent coefficients. Coefficients $c_z$ and $c_\gamma$ are real, but $c_w$  can be complex. As one can see, there is no direct interaction of the Chern-Simons (CS) vector boson $X_\mu$ with fields of the matter. It should be noted that Lagrangian \eqref{Lcs} does not contain term $\epsilon^{\mu\nu\lambda\rho} X_\mu A_\nu \partial_\lambda A_\rho$ which is not gauge invariant with respect to the electromagnetic $U(1)$ group.

The interest in the Chern-Simons model is due to the fact that the CS interaction is derived by an anomaly cancellation. 
This way is  theoretically attractive because  the contribution of extremely heavy fermions (not available for direct search at accelerators) to anomaly cancellation remains unsuppressed at low energies \cite{DHoker:1984izu,DHoker:1984mif}. 
These heavy fermions can act as mediators and  induce  observable interaction of the SM particles with new light vector particles from the "hidden" sector. A detailed  explanation of the origin  of the CS interactions \eqref{Lcs}, including using of a  toy UV model, can be found in \cite{Antoniadis:2009ze}.
The Chern-Simons interactions appear in various theoretical models, including extra dimensions and the string theory, see \cite{Antoniadis:2000ena,Coriano:2005own,Coriano:2007xg,Anastasopoulos:2006cz,Anastasopoulos:2008jt,Harvey:2007ca,Dudas:2009uq,Kumar:2007zza,Dror:2017ehi,Dror:2017nsg}. In addition to the need to study the CS theory as one of the possible extensions of the SM, the presence of the Levi-Civita symbol indicates the possible effects of violation of CP invariance in the CS theory, and perhaps the CS boson will be useful in solving some of the SM problems. On the other hand, if the CS boson is discovered, this will unequivocally indicate the presence of new physics at high energy scales. It is important to note that the CS boson was included in the proposal of the SHiP experiment \cite{Alekhin:2015byh}.

In order to start looking for the CS boson in experiments, one needs to know the main channels of the CS boson production and decay. 
In this paper we will  consider only the production of GeV-scale CS bosons. We will limit ourselves only to the case of the CS boson production in the decays of different types of mesons.
To do this, we will consider loop interaction of the CS boson with the SM fermions
and construct the effective Lagrangian of the CS bosons interaction with two different quarks. This will allow us to find more effective channels for the production of CS bosons and bring us closer to the experimental search for CS bosons.

The paper is organized as follows: in section~\ref{sec:1} we discuss theoretical aspects of the Chern-Simons model. In section~\ref{sec:2} we obtain the effective Lagrangian of the CS boson interaction with different quarks; in section~\ref{sec:3} we consider the CS boson production in mesons decays.
The summary and final discussion are given in section~\ref{sec:4}.  
Necessary technical clarifications and details about form-factors used are given in appendices.

\section{Chern-Simons model}
\label{sec:1}

The simplest case of the SM extension with non-trivial anomaly cancellation that involves the electromagnetic $U_{EM}(1)$ gauge group requires  small parameters of the photon mass or millicharge of new particles, see \cite{Boyarsky:2005hs,Boyarsky:2005eq,Antoniadis:2006wp, Antoniadis:2007sp}. These parameters are strongly restricted, see e.g. \cite{Williams:1971ms,Tu:2005ge,Davidson:1991si,Davidson:2000hf},  and suppress any visible effects. We get a similar situation when considering  non-trivial anomaly cancellation in the electroweak sector  of the SM. The small parameter, in this case, is the  sum of the electric charges of the electron and the proton, which is also limited to a very small value \cite{Marinelli:1983nd,Zyla:2020zbs}.

To avoid the need to deal with very small parameters of the models that suppress any visible effects  it is interesting to consider extension of the SM by an additional gauge group, see e.g. \cite{Hewett:1988xc} and references therein. In particular, Lagrangian \eqref{Lcs}  can be obtained in $U_X(1)$ gauge field  extension of the SM to the 
general theory with $SU_C(3)\times SU_W(2)\times U_Y(1)\times U_X(1)$ symmetry. One should also add to the theory heavy new chiral fermions that interact both with the gauge field of $U_X(1)$ and with the SM gauge fields. Herewith the SM fermions
are not charged with respect to the $U_X(1)$ group \cite{Antoniadis:2007sp}.

Lagrangian \eqref{Lcs} has $U_{EM}(1)$ gauge invariance, but unfortunately we need at least 6-dimension operators \cite{Alekhin:2015byh,Antoniadis:2009ze} to restore $U_Y(1)\times SU_W(2)$ invariance: 
\begin{align}
        \mathcal{L}_1&=\frac{C_Y}{\Lambda_Y^2}\cdot X_\mu (\mathfrak D_\nu H)^\dagger H B_{\lambda\rho} \cdot\epsilon^{\mu\nu\lambda\rho}+h.c.,\label{L1} \\
          \mathcal{L}_2&=\frac{C_{SU(2)}}{\Lambda_{SU(2)}^2}\cdot X_\mu (\mathfrak D_\nu H)^\dagger F_{\lambda\rho} H\cdot\epsilon^{\mu\nu\lambda\rho}+h.c.,\label{L2}  
\end{align}
where $\Lambda_Y$, $\Lambda_{SU(2)}$ are new scales of the theory; $C_Y$, $C_{SU(2)}$ are new dimensionless coupling constants; $H$ -- scalar field of the Higgs doublet; $B_{\mu\nu}=\partial_\mu B_\nu-\partial_\nu B_\mu$, $F_{\mu\nu}={\displaystyle -ig\sum_{i=1}^3\frac{\tau^i}{2} V^i_{\mu\nu} }$ -- field strength tensors of the $U_Y(1)$ and $SU_W(2)$ gauge fields. 

It is convenient to rewrite the coefficients before operators of dimension-6 as
 $C_Y/\Lambda_Y^2=C_1/v^2$ and $C_{SU(2)}/\Lambda_{SU(2)}^2=C_2/v^2$, where $C_1=c_1+{\rm i} c_{1{\rm i}}$ and  $C_2=c_2+{\rm i} c_{2{\rm i}}$ are dimensionless coefficients, $v$ is the vacuum expectation value of the Higgs field. In this case Lagrangians \eqref{L1}, \eqref{L2} after the electroweak symmetry breaking generate Lagrangian of three fields interactions \eqref{Lcs}   with coefficients in unitary gauge:
\begin{align}
&    c_z=  -c_{1{\rm i}}g' + \frac{c_{2}}2 g^2, \label{L7}\\
&    c_\gamma = c_{1{\rm i}} g + \frac{c_{2}}2 g g^{'} , \label{L8}\\
&    c_w= \frac{c_2+{\rm i} c_{2{\rm i}}}2 g^2\equiv \Theta_{W1}+{\rm i} \Theta_{W2}.\label{L9}
\end{align}

In \cite{Alekhin:2015byh} it was pointed out that from dimension-6 operators \eqref{L1}, \eqref{L2} follows $c_z/c_\gamma=\tan\theta_W$. However, we see from relations \eqref{L7} -- \eqref{L9} that ratio $c_z/c_\gamma$ depends on two unknown parameters $c_{1{\rm i}}$ and $c_2$.
It is evident that the ratio of these quantities has a simple form only in the particular case when the real part of the parameter $C_2$ is zero ($c_z/c_\gamma=-\tan\theta_W$), or the imaginary part of the parameter $C_1$ is zero ($c_z/c_\gamma=\cot\theta_W$). As one can see, relations \eqref{L7} -- \eqref{L9} allow us to write a relation between the real part  of parameter $C_2$ and $c_z$, $c_\gamma$ that gives $\Theta_{W1}=\cos^2\theta_W(c_z+\tan\theta_W c_\gamma)$. It is also interesting that the real part of parameter $C_1$  (parameter $c_1$) is not included into \eqref{L7} -- \eqref{L9} and the imaginary part of parameter $C_2$  (parameter $\Theta_{W2}$) is an independent parameter.

It should be noted that relations \eqref{L7} -- \eqref{L9} and corresponding conclusions are valid only in the assumptions that the Chern-Simons theory is derived only by dimension-6 operators \eqref{L1}, \eqref{L2}. 
But it may turn out that the main or a significant contribution to Lagrangian \eqref{Lcs} comes  from higher dimension operators. So, following \cite{Alekhin:2015byh},  below we will consider $c_z$, $c_\gamma$, $c_w$  as independent dimensionless parameters.

\begin{figure}[t]
\centering
\includegraphics[width=0.5\textwidth]{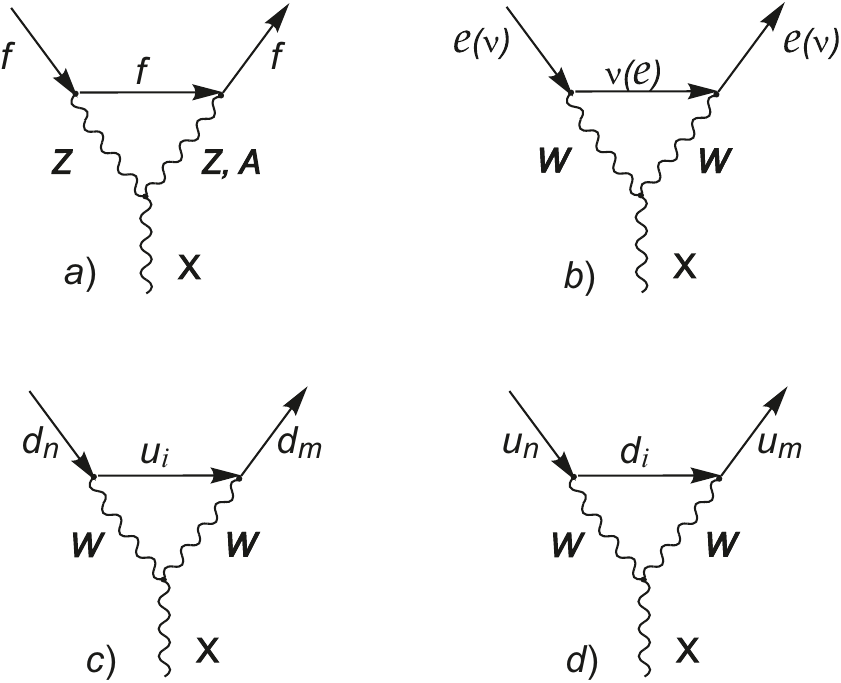} \caption{CS boson loop interactions with two fermions of the SM.}
\label{fig:XfermInt}
\end{figure}

\section{Effective interaction of the CS boson with different quarks}\label{sec:2}

Lagrangian \eqref{Lcs} gives diagrams of the CS boson loop interactions with two SM fermions presented in  figure~\ref{fig:XfermInt}. 
Since we want to consider the CS boson production  in decays of mesons we only need diagrams $(c)$ and $(d)$, where a heavy initial particle decays in a light particle with the production of the CS boson. It means we can consider in Lagrangian \eqref{Lcs} only the interaction of the CS boson with $W$ bosons and put $c_z=c_\gamma=0$ for simplicity.

\begin{figure}[t]
\centering
\includegraphics[width=0.25\textwidth]{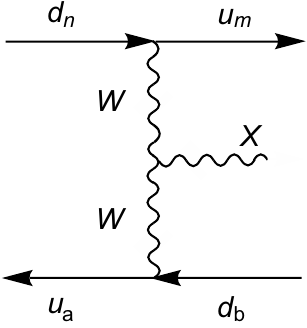} \caption{Interaction of the CS boson with four fermions.}
\label{fig:XfermInt4}
\end{figure}

Certainly, one can present  four fermion interaction with the CS boson without loop, see figure~\ref{fig:XfermInt4}, but this interaction will be suppressed by $G_F^2$ and, as it will be shown in section~\ref{sec:InterLagr}, can be neglected compared to the loop interaction of the CS boson with two fermions.

\subsection{Interaction of the CS boson with two different down-type quarks}\label{sec:i}

Computation of the loop diagram for the interaction of the CS boson with down-type quarks gives us, see figure  \ref{fig:Xbs} and appendix \ref{appA}, the following amplitude of the heavy down-type quark ($d_n$) decay into light down-type quark $(d_m)$  and the CS boson in the unitary gauge:
\begin{equation}\label{Mfim}
    M_{fi}=\frac{g^2}{2}\sum_{i=u,c,t} V_{d_m i}^+V_{id_n}\,\overline{d_m}(p^\prime)  Y^{\mu}_{(i)}  d_n(p) \epsilon^{*\lambda_X}_{\mu},
\end{equation}
where  $Y^{\mu}_{(i)}$ has  a rather cumbersome form
\begin{multline}\label{unitrep1m}
    Y^{\mu}_{(i)}  =
   \hat\Lambda_0 \left\{ \vphantom{\frac12}
  2 (x + y - 1)
   \left\{\Theta_{W1}  (2 y-1) -{\rm i}\Theta_{W2} \right\} -
     2 y
   \left\{\Theta_{W1} (1 - 2x-2 y) +{\rm i}\Theta_{W2}  \right\}-\right. \\ -
   \frac{x}{M_W^2}
      \left[c_w^* \left\{ (3x+3y-1)m^2 +y m^{\prime\, 2}-4y (p\, p^\prime)  \right\}-\right. \\  -2\Theta_{W1} \left\{(x+y)^2m^2+y^2m^{\prime\, 2}-2y(x+y)(p\, p^\prime)\right\}+\\
    \left. \left.  +2{\rm i}\Theta_{W2}  \left\{ (x+y)m^2-2y (p\, p^\prime) \right\}
      \right] \vphantom{\frac12}  \right\}  \, p^\prime_{ \lambda}\,\gamma_{ \rho}\,p_{ \nu}\,\hat P_L \epsilon^{{ {\mu}}{ {\nu}}{ {\lambda}}{ {\rho}}} +  \\
          +\hat\Lambda_0 \,
      m^\prime\, y
   \left\{\Theta_{W1}( (1 - 2x-2 y) p + (2 y-1) p^\prime )_{ \lambda}+{\rm i}\Theta_{W2} (p-p^\prime)_{ \lambda} \right\} \gamma_{ \rho}\gamma_{ \nu} \hat P_L \, \epsilon^{{ {\mu}}{ {\nu}}{ {\lambda}}{ {\rho}}}+\\
      + \hat\Lambda_1 \left\{ -{\rm i}\Theta_{W1}\gamma_{ {\rho}}\gamma_{ \lambda}\gamma_{ {\nu}}-2{\rm i}  \frac{p^\prime_{ \lambda} \gamma_{ \rho}\, p_{ \nu}}{M_W^2} 
      \Theta_{W1}(1-3x)  \right\} \hat P_L
      \epsilon^{{ {\mu}}{ {\nu}}{ {\lambda}}{ {\rho}}}+\\
      +\hat\Lambda_0 \left\{ \vphantom{\frac12}
    -(x + y - 1) m
   \left\{\Theta_{W1}[ (1 - 2x-2 y) p + (2 y-1) p^\prime ]_{ \lambda}+{\rm i}\Theta_{W2} (p-p^\prime)_{ \lambda} \right\} \gamma_{ \rho}\gamma_{ \nu}\right.\\ \left.
   + \frac{x\, m m^\prime }{M_W^2}
      \left[\Theta_{W1} (1-x)   + {\rm i}\Theta_{W2}  (2y+x-1)
      \right] \,p^\prime_{ \lambda}\, \gamma_{ \rho} \, p_{ \nu} \right\} \hat P_R \, \epsilon^{{ {\mu}}{ {\nu}}{ {\lambda}}{ {\rho}}}-\\
       - \hat\Lambda_1 \,{\rm i}  (p-p^\prime)_{ \lambda} \gamma_{ \rho}\gamma_{ \nu} \frac{c_w  m \hat P_R-c_w^*  m^\prime \hat P_L}{2M_W^2} 
      \epsilon^{{ {\mu}}{ {\nu}}{ {\lambda}}{ {\rho}}},
\end{multline}
integral operators $\hat \Lambda_{0(1)}$ are defined as
\begin{align}
    & \hat \Lambda_0 ={\rm i} \frac{\pi^2}{(2\pi)^4} \int_0^{1}{\rm d} x\int_0^{1-x}{\rm d} y\, \frac{1}{D(m_i)},\label{Lambda0m} \\
     & \hat \Lambda_1 = -\frac{\pi^2}{(2\pi)^4} \int_0^{1}{\rm d} x\int_0^{1-x}{\rm d} y\, \ln\frac{\Lambda^2 x}{D(m_i)},\label{Lambda1m}
\end{align}
here
\begin{equation}\label{D1m}
    D(m_i)=x m_i^2+(1-x)M_W^2+xy(M_X^2+m^2-m^{\prime\, 2})-x(1-x)m^2-y(1-y)M_X^2,
\end{equation}
$\hat P_{R(L)}$ -- projection operators on the right(left)-handed
chirality states, $p$ is the 4-momentum and $m$ is the mass of the initial down-type quark $(d_n)$, $p^\prime$  is the 4-momentum and $m^\prime$ is the mass of the final down-type quark $(d_m)$, $m_i$ is the mass of the virtual up-type $u_i$ quark,
$M_W$ and $M_X$ are the masses of the $W$ and CS vector bosons, $\Lambda$ is the regularization parameter that should be set to infinity.

The divergent part of $Y^{\mu}_{(i)}$ is hidden in the operator $\hat\Lambda_1$ \eqref{Lambda1m}. We can extract in this operator singular and finite parts:
\begin{equation}\label{Lambda1regm}
    \hat \Lambda_1 =  \hat \Lambda_1^{sing}+\hat \Lambda_1^{fin} =-\frac{\pi^2}{(2\pi)^4} \int_0^{1}{\rm d} x\int_0^{1-x}{\rm d} y\,\left\{ \ln\frac{\Lambda^2 x}{\mu^2}-
    \ln\frac{D(m_i)}{\mu^2 }\right\},
\end{equation}
where $\mu$ is some parameter with the dimension of mass.  
It should be noted that operator $\Lambda_1^{sing}$ does not depend on the mass of the virtual quark in the loop ($m_i$) and it acts on an expression that does not depend on $m_i$ too. So, after summation over the all virtual quarks in  \eqref{Mfim}
we get a term proportional to
$$  \sum_{i=u,c,t} V_{d_m i}^+V_{id_n} \ln\frac{\Lambda^2 x}{M_W^2}=(V^+V)_{d_m d_n} \ln\frac{\Lambda^2 x}{M_W^2}  =0, \quad m \neq n,$$
due to the unitarity of the CKM-matrix in the SM.
So, in the expression \eqref{unitrep1m} we  can replace operator $\hat\Lambda_1$ by finite operator  $\hat\Lambda_1^{fin}$
\begin{equation}\label{Lambda1finm}
    \hat \Lambda_1^{fin} = \frac{\pi^2}{(2\pi)^4} \int_0^{1}{\rm d} x\int_0^{1-x}{\rm d} y \ln\frac{D(m_i)}{\mu^2}.
\end{equation}
With the help of the unitary condition for the CKM-matrix, it is not difficult to  show that the amplitude of the process does not  depend on the value of the parameter $\mu$, see appendix \ref{appA}. 
In the following, we will put   $\mu=M_W$  for the convenience of computations. 
It should be noted that the established fact of the divergent part of the CS boson loop interaction with quarks of two different flavors being zero (as a consequence of the CKM matrix unitarity in the SM) is consistent with the results of  \cite{Dror:2017ehi,Dror:2017nsg}.

An explicit form of the $Y^{\mu}_{(i)}$ can be obtained after applying the operators  $\hat \Lambda_0$, $\hat \Lambda_1^{fin}$ and performing summation over the virtual up-type quarks in \eqref{unitrep1m}, see appendix \ref{appB}. 
The amplitude of a heavy down-type quark ($d_n$) (with mass $m$ and 4-momentum $p$) decay into a light down-type quark $(d_m)$ (with mass $m^\prime$ and 4-momentum $p^\prime$)  and the CS boson in the unitary gauge \eqref{Mfim} is convenient to present in the form
\begin{equation}\label{Mfi1m}
    M_{fi}=\frac{g^2}{32\pi^2}\,  \frac{m_t^2}{M_W^2}\, V_{d_m t}^+V_{t d_n}  \,\overline{d_m}(p^\prime)\, J^{\mu,d}(p,p^\prime)\,  d_n(p) \epsilon^{*\lambda_X}_{\mu} ,
\end{equation}
where $m_t$ is mass of the top quark and
\begin{align}
& J^{ \mu}= \left(a_L^d\,\hat
 P_R\,\gamma_{ \rho}\, \gamma_{ \lambda}\,\gamma_{ \nu}\,\hat
 P_L +\right. \nonumber \\
& 
 b_L^d \,\frac{p^\prime_{ \lambda}\,p_{ \nu}}{M_W^2}\,\hat
 P_R\,\gamma_{ \rho}\,\hat
 P_L  + c_L^d\, \frac{m^\prime\, p^\prime_{ \lambda}}{M_W^2}\,\hat
 P_L\,\gamma_{ \rho}\,\gamma_{ \nu}\,\hat
 P_L  +d_L^d\,  \frac{m^\prime\, p_{ \lambda}}{M_W^2}\,\hat
 P_L\,\gamma_{ \rho}\,\gamma_{ \nu}\,\hat
 P_L  + \nonumber \\ 
 & \left. b_R^d \, \frac{m m^\prime}{M_W^2}\,\frac{p^\prime_{ \lambda}\,p_{ \nu}}{M_W^2}\,\hat
 P_L\, \gamma_{ \rho}\,\hat
 P_R  +c_R^d\,\frac{m\,  p^\prime_{ \lambda}}{M_W^2}\,\hat
 P_R\,\gamma_{ \rho}\,\gamma_{ \nu}\,\hat
 P_R + d_R^d\, \frac{m\,  p_{ \lambda}}{M_W^2}\,\hat
 P_R\,\gamma_{ \rho}\,\gamma_{ \nu}\,\hat
 P_R \right)\,\epsilon^{{ {\mu}}{ {\nu}}{ {\lambda}}{ {\rho}}}.\label{Mfi2m}
\end{align}

Coefficients in \eqref{Mfi2m} 
in the first approximation can be considered independent of the masses of down-type quarks. They can be computed numerically, but we also managed to find the analytical expression for these coefficients with less than $1\%$ difference from their values  obtained numerically, see appendix \ref{appB}. 
The values of the coefficients are given in table  \ref{tab:coefupdownm}. Coefficients at $\Theta_{W1}$ are imaginary with sufficient accuracy, but coefficients at $\Theta_{W2}$ are real.
Taking into account that near the coefficients $b^d_{L,R}$, $c^d_{L,R}$, $d^d_{L,R}$ there are suppressing factors, one can see that the main contribution to \eqref{Mfi2m} comes from the term $a^d_L$ (if $\Theta_{W1}\neq 0$). It should be noted that coefficient  $a_L^d$ depends only on one parameter $(\Theta_{W1})$ unlike almost all other coefficients depending on both parameters: $\Theta_{W1}$ and $\Theta_{W2}$.

\subsection{Interaction of the CS boson with two different up-type quarks}

Let us now consider the interaction of the CS boson with up-type quarks. The amplitude of a heavy up-type quark ($u_n$) decay into a light up-type quark $(u_m)$  and the CS boson in the unitary gauge is given by
\begin{equation}\label{Mfimup}
    M_{fi}=\frac{g^2}{2}\sum_{i=d,s,b} V_{i u_m }^+ V_{u_n i}\,\overline{u_m}(p^\prime)  Y^{\mu}_{(i)}  u_n(p) \epsilon^{*\lambda_X}_{\mu},
\end{equation}
where $Y^{\mu}_{(i)}$ is defined in \eqref{unitrep1m}. However, in this case $p$ is the 4-momentum and $m$ is mass of the initial up-type quark $(u_n)$, $p^\prime$  is the 4-momentum and $m^\prime$ is mass of the final up-type quark $(u_m)$, $m_i$ is mass of the virtual down-type quark ($d_i$).

\begin{table}[t]
\centering
\begin{tabular}{|c|l|c|l|}
 \hline
 coeff. & value  & coef. & value$\cdot10^{6}$ \\
 \hline
 $a_L^d$ & $- 0.13 \,{\rm i}\, \Theta_{W1}$ & $a_L^{up}$  & $(-0.98 + 1.13 {\rm i}) \Theta_{W1}$ \\ \hline
  $b_L^d$ & $0.05 {\rm i} \Theta_{W1}+ 4\cdot 10^{-5} \Theta_{W2}$ & $b_L^{up}$  & $(-6.4 + 7.4 {\rm i})\Theta_{W1} - (1.1 +
    0.98 {\rm i}) \Theta_{W2}$  \\ \hline
  $c_L^d$ & $-0.058\, {\rm i}\, \Theta_{W1} - 0.098 \,\Theta_{W2}$ & $c_L^{up}$  & $(0.27 - 0.31 {\rm i}) \Theta_{W1} + (0.57 +
    0.49 {\rm i}) \Theta_{W2}$ \\ \hline
  $d_L^d$ & $0.086\, {\rm i} \,\Theta_{W1} + 0.098 \,\Theta_{W2}$ & $d_L^{up}$  & $(0.38 - 0.44 {\rm i}) \Theta_{W1} - (0.57 +
    0.49 {\rm i}) \Theta_{W2}$  \\ \hline
  $b_R^d$ & $-0.028\,  {\rm i} \,\Theta_{W1}$ & $b_R^{up}$  & $(-0.64 + 0.75 {\rm i}) \Theta_{W1} $
  \\ \hline
  $c_R^d$ & $0.086\, {\rm i}\, \Theta_{W1} - 0.098\, \Theta_{W2}$ & $c_R^{up}$  & $(0.87 -  {\rm i}) \theta_{W1} + (1.1 +
    0.98 {\rm i}) \Theta_{W2}$  \\ \hline
  $d_R^d$ & $-0.058 \,{\rm i}\, \Theta_{W1} + 0.098\, \Theta_{W2}$ & $d_R^{up}$  & $(-0.23+ 0.25 {\rm i}) \Theta_{W1} - (1.1 +
    0.98 {\rm i}) \Theta_{W2}$ \\ \hline
\end{tabular}
    \caption{Coefficients in the expression for the amplitude of a heavy  quark   decay into a down-type quark and the CS boson  in \eqref{Mfi2m}. Superscript $d$ stands for the decay of down-type quark $d_n\rightarrow d_m +X$, superscript $up$ is for the decay of up-type quark $c \rightarrow u +X$.}
    \label{tab:coefupdownm}
\end{table}

It should be noted that the mass of the virtual down-type quark ($d_i$) in $Y^{\mu}_{(i)}$ is included only in function $D(m_i)$ \eqref{D1m}, but for all virtual down-type quarks $m_i \ll M_W$.
So, in the first approximation function, $D(m_i)$ can be considered independent of the mass of virtual down-type quark $m_i$ and can be taken as $D(m_i)=(1-x)M_W^2$. It means that $Y^{\mu}_{(i)}\approx Y^{\mu}$ and
\begin{equation}
    \sum_{i=d,s,b} V_{u_m i}^+V_{iu_n}\,  Y^{\mu}_{(i)} \approx Y^{\mu} \sum_{i=d,s,b} V_{u_m i}^+V_{iu_n}=0, \quad m \neq n. 
\end{equation}
Thus, in the first approximation the amplitude of a heavy up-type quark ($u_n$) decay into a light up-type quark $(u_m)$  and the CS boson is zero.  
If we perform accurate calculations and present the amplitude in a form similar to the case of the down-type quarks:
\begin{equation}\label{Mfi1mup}
    M_{fi}=\frac{g^2}{32\pi^2}\,  \frac{m_t^2}{M_W^2}\, V_{s t}^+V_{t b}  \,\overline{u_m}(p^\prime)  J^{\mu,up}  u_n(p) \epsilon^{*\lambda_X}_{\mu} ,
\end{equation}
where $J^{\mu,up}$ has form
\eqref{Mfi2m}, but with coefficients with superscript $up$, we get values presented in the table  \ref{tab:coefupdownm}.

\subsection{Lagrangian of the effective interaction of the CS boson with two different quarks}\label{sec:InterLagr}

After analyzing the data in table  \ref{tab:coefupdownm} one can conclude that the production of CS bosons in decays of up-type quarks is substantially suppressed in comparison with the production of CS bosons in decays of down-type quarks.  

If parameter $\Theta_{W1}$ is nonzero, then the main contribution for the CS boson production from down-type quarks comes from the term $a^d_L=-{\rm i} a\, \Theta_{W1}$, see \eqref{Mfi2m}.
Using relation 
\begin{equation}\label{gggepsm}
    \epsilon^{\alpha\mu\nu\rho}\gamma_\mu \gamma_\nu \gamma_\rho= 6{\rm i} \gamma^\alpha
    \gamma^5,
\end{equation}
 one can write the effective Lagrangian of the GeV-scale CS boson interaction with different down-type quarks in the form
\begin{equation}\label{Mfi5m}
    \mathcal{L}^{CS}_{quarks}=  \sum_{m< n}\Theta_{W1}\left( C_{mn}\, \overline{d_m}\, \gamma^{\mu}\,\hat
 P_L  \,  d_n X_{\mu}+C_{nm}^+\, \overline{d_n}\, \gamma^{\mu}\,\hat
 P_L  \,  d_m X_{\mu}\right),
\end{equation}
where the summation occurs over the generations of quarks,
\begin{equation}\label{Cdm}
    C_{mn}= \frac{3a}{2\sqrt{2}\pi^2}\, G_F m_t^2\,V_{d_m t}^+V_{t d_n}
\end{equation}
and
\begin{equation}
  |C_{sb}| = 1.97\cdot 10^{-4},\quad
  |C_{db}|=   4.43\cdot 10^{-5},\quad
   |C_{ds}|=   1.77\cdot 10^{-6}.
\end{equation}
Interaction of the GeV-scale CS boson  with up-type quarks can be neglected. These results are consistent with the results of  \cite{Dror:2017ehi,Dror:2017nsg}.

There are two points to pay attention to. First, Lagrangian \eqref{Mfi5m} is valid only for  the GeV-scale CS boson interaction with different quarks. The case of the CS boson interaction with the same quarks  must be considered separately. In the last case, we can not eliminate divergences in the loop integral using only the condition of the  CKM-matrix unitarity. Second, Lagrangian \eqref{Mfi5m} is slightly  similar to the interaction Lagrangian of quarks with $W$ bosons in the SM. As in the case of the SM the term $\overline{d_m}\, \gamma^{\mu}\,\hat
 P_L  \,  d_n X_{\mu}$ has no symmetry under the separate  action  of  charge conjugation ($\hat C$) and  parity transformation ($\hat P$) operators. But this term will have symmetry under the simultaneous action of charge conjugation and parity transformation operators
 \begin{equation}\label{CPtransf}
\hat C \hat P \overline{d_m}\, \gamma^{\mu}\,\hat
 P_L  \,  d_n X_{\mu}= \overline{d_n}\, \gamma^{\mu}\,\hat
 P_L  \,  d_m X_{\mu},
\end{equation}
 if we impose a reasonable condition 
  \begin{equation}\label{CPtX}
\hat C \hat P X_\mu= - X^\mu.
\end{equation}
Thus, having analyzed the effective Lagrangian \eqref{Mfi5m}, we come to the conclusion that the CS boson has no well-defined symmetry under the separate action of  charge conjugation ($\hat C$) and  parity transformation ($\hat P$) operators, but is even under the simultaneous action of $\hat C \hat P $ transformation like $Z$ boson. 

As in the case of the  interaction Lagrangian of quarks with $W$ bosons in the SM, the effective Lagrangian \eqref{Mfi5m} will be $\hat C \hat P $  invariant only if matrix 
$C_{mn}$ is real.

\section{The CS boson production in decays of mesons}\label{sec:3}

Since the CS boson interaction with up-type quarks is strongly suppressed, we will consider only production of CS bosons from mesons containing $b$ or $s$ quarks. Such lightest mesons are $B$ mesons and $K^\pm$, $K^0_S$, $K^0_L$ mesons.

Dominant  reactions of the $B$ meson decay with the CS boson production are  two-body decays into pseudoscalar mesons ($K$ and  $\pi$ mesons); scalar mesons $K^{0*}(700)$ and $K^{0*}(1430)$; vector mesons $K^*(892)$, $K^*(1410)$, $K^*(1680)$; pseudovector mesons $K_1(1270)$ and $K_1(1400)$;  tensor final meson state $K_2(1430)$.

For the initial kaons states, the only possible 2-body decay with the CS boson production  is the process
\begin{equation}
    K \to \pi +X.
\end{equation}
There are 3 types of kaons -- $K^{\pm}, K^{0}_{L}, K^{0}_{S}$. Although the decay width for each of them is  given by the same loop factor ($ C^d_{sd}$) the branching ratios differ. The first reason is that these kaons have different decay widths. The second reason is that $\pi^0$ meson is the $CP$-odd eigenstate, $K^{0}_{S}$ is approximately the $CP$-even eigenstate, and $K^{0}_{L}$  is approximately the $CP$-odd eigenstate. 
The $CP$ parity of the final state in the reaction  $K^{0}_{L}\to \pi^0 X$ is $(-1)^L(CP)_{\pi^0}(CP)_X=+1$ (since the total angular momentum of the initial meson is zero, the final particles must have orbital angular momentum $L=1$). Therefore, the decay width of the reaction $K^{0}_{L}\to \pi^0 X$ is proportional to the CKM $CP$-violating phase \cite{Gunion:1989we,Leutwyler:1989xj}.

Branching of the two-body meson decay $h\rightarrow h' + X$ is defined as
\begin{equation}\label{brdef}
Br(h\rightarrow h^\prime X)=\frac{1}{\Gamma_h}\,\frac{\left|M_{h\rightarrow h^\prime X}\right|^2}{8\pi M_h^2 }\, |\vec k|,
\end{equation}
where
\begin{equation}\label{l2.32a2}
|\vec k|=\frac{\sqrt{\lambda(M_{h}^2,M_{h^\prime}^2,M_X^2)}}{2M_{h}}
\end{equation}
and 
$\lambda (x,y,z) = x^2 + y^2 + z^2 - 2xy - 2yz - 2zx
$ is the K\"all\'en function \cite{Kallen:1964lxa}.

The amplitude of the decay of a containing $d_n$ quark $h$ meson into a containing $d_m$ quark $h^\prime$ meson and the CS boson has form, see \eqref{Mfi5m},
\begin{equation}\label{amplitude}
    M_{h\rightarrow h^\prime X}=\Theta_{1W}   C_{mn} \, \langle h^\prime(p^\prime)|\bar d_m\gamma^{\mu} \hat P_L d_n|h(p)\rangle\, \epsilon^{*\lambda_X}_{\mu},\\
\end{equation}
where the absolute  values of $  C_{mn} $ are given by \eqref{Cdm}. 
In the case of decay of  $K^0_S$, $K^0_L$ mesons, one has to be more careful due to the presence of a certain CP parity in them, namely
\begin{align}
    M_{K^0_L\rightarrow \pi^0 X}=\Theta_{1W}  Im[ C_{ds}] \, \langle \pi^0(p^\prime))|\bar d\gamma^{\mu} \hat P_L s|K^0_S(p)\rangle\, \epsilon^{*\lambda_X}_{\mu},\label{amplitudeKS}\\
    M_{K^0_S\rightarrow \pi^0 X}=\Theta_{1W}  Re[ C_{ds}] \, \langle \pi^0(p^\prime))|\bar d\gamma^{\mu} \hat P_L s|K^0_L(p)\rangle\, \epsilon^{*\lambda_X}_{\mu},\label{amplitudeKL}
\end{align}
where $Re[ C_{ds}] =-1.62\cdot10^{-6}$ and $Im[ C_{ds}]=-6.69\cdot 10^{-7}  $.

The average over meson states  $\langle h'(p'))|\bar d_m\gamma^{\mu} \hat P_L d_n|h(p)\rangle$ can be obtained with help of formalism summarized e.g.  in \cite{Boiarska:2019jym}.

In the rest frame of the initial $h$ meson we have 
$ p=(M_h,0)$. 
It decays into the $h'$ meson and the CS boson with momentums
\begin{equation}\label{p'pX}
    p^\prime=(E^\prime,-\vec k), \quad p_X=(E_X,\vec k),\quad p_X=p-p^\prime.
\end{equation}
where direction of spatial vector $\vec k$ is arbitrary, but module $|\vec k|$ is defined by \eqref{l2.32a2}.
Let's guide Z-axis along the spatial momentum of the CS boson, then 
4-vector of the CS boson polarisation is defined as
\begin{equation}\label{polaris}
    \epsilon^{(\pm)}_X=\left(0, 1,\mp {\rm i},0\right)/\sqrt{2},\quad
    \epsilon^{(0)}_X=\left(|\vec k|, 0,0,E_X\right)/M_X,
\end{equation}
where $\epsilon^{(\pm)}_X$ corresponds to the spin projection $\pm1$ and $\epsilon^{(0)}_X$ corresponds to  zero spin projection on the momentum direction. 

If  $\langle h^\prime(p^\prime))|\bar d_n\gamma^{\mu} \hat P_L d_m|h(p)\rangle$ depends only on $p^\mu$ or $p^{\prime\,\mu}$, then only the longitudinal component of the CS boson polarisation gives contribution into the amplitude of the reaction because   $\epsilon^{(\pm)}_X p=\epsilon^{(\pm)}_X p'=0$.  It will be useful to write also
\begin{equation}\label{epspp}
 \epsilon^{(0)}_X \,p =\epsilon^{(0)}_X \,p^\prime= |\vec k|\frac{M_h}{M_X},\quad  \epsilon^{(0)}_X \,p_X =0.
\end{equation}

\begin{table}[t]
\centering
\resizebox{\textwidth}{!}{%
\begin{tabular}{|l|c|c|c|c|}
 \hline
 Process& final meson  & ${\displaystyle\lim_{m_X\rightarrow 0}\left(\frac{m_X}{1\text{GeV}}\right)^2\frac{\text{Br}(m_X)}{\theta_{W1}^2}}$ & Closing mass [GeV] & appendix \\
 \hline
  $K^\pm\to X\pi^{\pm}$ & pseudo scalar &$2.49\cdot10^1$  & 0.35  & \ref{app:sameparity} \\ \hline
  $K^0_L\to X\pi^0$ & pseudo scalar &$1.56\cdot 10^1$ & 0.36 & \ref{app:sameparity} \\ \hline
  $K^0_S\to X\pi^0$ & pseudo scalar &$\,\,1.61\cdot 10^{-1}$  & 0.36& \ref{app:sameparity} \\ \hline
  $B^{\pm} \to X \pi^{\pm}$&pseudo scalar &  $2.37\cdot10^2$ &  5.14 & \ref{app:sameparity} \\ \hline
  $B^{\pm} \to X K^{\pm}$& pseudo scalar & $7.73\cdot 10^3$  & 4.79 & \ref{app:sameparity} \\ \hline
  $B^{\pm} \to X K^{*,\pm}_{0}(700)$ & scalar  &$ 1.43\cdot 10^4 $ & 4.46  & \ref{app:anotherparity} \\ \hline
  $B^{\pm} \to X K^{*,\pm}(892)$& vector & $9.14 \cdot 10^3$ & 4.39 & \ref{app:vector} \\ \hline
  $B^{\pm} \to X K^{\pm}_{1}(1270)$ & pseudo vector & $1.72\cdot 10^4$ & 4.01 & \ref{app:pseudovector} \\ \hline
  $B^{\pm} \to X K^{\pm}_{1}(1400)$&  pseudo vector & $2.34\cdot 10^2$ & 3.88 & \ref{app:pseudovector} \\ \hline
  $B^{\pm} \to X K^{*,\pm}(1410)$& vector  &  $3.99\cdot 10^3$ & 3.86 & \ref{app:vector} \\ \hline
  $B^{\pm} \to X K^{*,\pm}_{0}(1430)$& scalar  & $1.85\cdot 10^3$ & 3.85 & \ref{app:anotherparity} \\ \hline
  $B^{\pm} \to X K^{*,\pm}_{2}(1430)$& tensor  & $6.03\cdot 10^3$  & 3.85 & \ref{app:tensor} \\ \hline
  $B^{\pm} \to X K^{*,\pm}(1680)$&  vector & $2.53\cdot 10^3$  & 3.56 & \ref{app:vector} \\ \hline
\end{tabular}}
    \caption{Properties of the main production channels of the CS boson from kaons and $B$ mesons. \textit{First column}: decay channels; \textit{Second column}: type of final mesons; \textit{Third column}: branching ratios of 2-body meson decays evaluated at $m_{X}\rightarrow 0$  and normalized by $\theta_{W1}^2$. For $B$ mesons the numerical values are given for $B^{\pm}$ mesons; in the case of $B^0$ meson all the given branching ratios should be multiplied by a factor of $0.93$ that comes from the difference in total decay widths of $B^{\pm}$ and $B^0$ mesons~\protect\cite{Zyla:2020zbs}; \textit{Fourth column}: closing mass, i.e. the difference between the masses of the initial and the final mesons; \textit{Fifth column}: a reference to the appendix with details about form-factors used.}
    \label{tab:BR}
\end{table}

Results for the branchings for the corresponding reactions  are presented in table \ref{tab:BR}, where for the decays of the $B$ mesons we took into account only the lightest final excited $K$ meson states from its scalar, pseudoscalar, vector, pseudovector and tensor states. 
The branching dependencies on the CS boson mass for decays of charged $K$ mesons and neutral $K^0_L$, $K^0_S$ mesons are shown in figure~\ref{fig:KBX}(a).  
The branching dependencies on the CS boson mass for decays of charged $B$ mesons are shown in figure~\ref{fig:KBX}b, where, for clarity, we have summed up the contributions over the (pseudo)scalar, (pseudo)vector and tensor channels.
Because of the inverse quadratic 
divergence of the amplitude of meson decay at small masses of the CS boson (it originates from its longitudinal component), values of the reaction branchings are presented for quantity   ${\displaystyle\lim_{m_X\rightarrow 0}\left(\frac{m_X}{1\text{GeV}}\right)^2\frac{Br(m_X)}{\theta_{W1}^2}}$.
Main contribution to the uncertainties in the presents results  follows from uncertainties in meson transition form-factors. Our results are consistent with the results of \cite{Dror:2017nsg}, where some of the reactions we considered had been calculated.

\section{Summary}\label{sec:4}

The goal of our study was to describe the production of the new GeV-scale vector particle $X_\mu$ (Chern-Simons boson) in the decays of mesons starting from the effective interaction of this boson with electroweak gauge bosons \eqref{Lcs}. 

We consider the loop interaction of the CS boson with quarks of different flavors through $W$ bosons. In this case, see figure~\ref{fig:XfermInt}(c) and figure~\ref{fig:XfermInt}(d), we have the CS boson interaction with two down-type or two up-type quarks.  We have shown that the divergent terms in the calculation of loop diagrams are automatically canceled out due to the unitarity of the CKM matrix, see appendix \ref{appA}. Unfortunately, this result is valid only for the CS boson interaction with quarks of different flavors. The case of the interaction of the CS boson with two identical quarks or with leptons requires a separate, more accurate consideration and some renormalization scheme, which is the subject of further research.

We have shown that the interaction of
the CS boson with up-type quarks is sufficiently suppressed compared  with down-type quarks. This is due to the fact that when the up-type quarks interact with the CS boson, the masses of the virtual down-type quarks in the loop are small compared to the $W$ boson mass. As a result, in the first approximation, there is no dependence of the loop terms on the mass of virtual down-type quarks and interaction of up-type quarks is suppressed by the condition of the unitarity of the CKM matrix.

\begin{figure}[t]
\centering
\includegraphics[width=\textwidth]{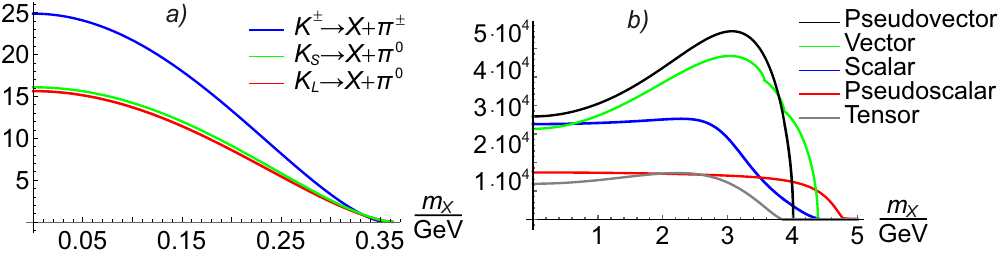} \caption{Dependence of the branching $Br = \, {\displaystyle\theta_{W1}^{-2}\left(\frac{m_X}{1\text{GeV}}\right)^2Br(h\rightarrow h^\prime X)}$ of reactions of the CS boson production  on the CS boson mass: a -- reaction of the CS boson production in  two-body decays of charged $K$ mesons and neutral $K^0_L$, $K^0_S$ mesons (values of the branching for the reaction $K^0_S \rightarrow \pi^0 + X$ are multiplied by $10^2$); b -- reaction of the CS boson production in two-body decays of charged $B$ mesons (contributions over the (pseudo)scalar, (pseudo)vector and tensor channels of the $B$ meson decays are summed up).}
\label{fig:KBX}
\end{figure}\vspace{-0.5em}

We construct the effective Lagrangian of the CS boson interaction  with two
different down-type quarks \eqref{Mfi5m}. It turned out that the effective Lagrangian depends only on one of
two  unknown couplings ($\Theta_{W1}$) of the CS boson interaction with $W$ boson, see  \eqref{Lcs}. It should be noted that in the interaction of the CS boson with quarks \eqref{Mfi5m}, the CS boson behaves like a CP even particle, see \eqref{CPtX}.\vspace{-0.5em}

We consider the production of the CS boson in  decays of mesons. Since CS bosons interact effectively  only with down-type quarks,  they can be produced in decays of mesons containing  $b$ or $s$ quarks. Such lightest mesons 
are $B$ mesons and $K^\pm$ mesons. Due to the contribution of longitudinal polarization 
  of the CS boson, the decay width of the reactions increases with decreasing boson mass as $M_X^{-2}$. So, for convenience, we computed quantity  ${\displaystyle \theta_{W1}^{-2}\left(\frac{m_X}{1\text{GeV}}\right)^2 Br(h\rightarrow h^\prime X)}$.

We consider the production of the CS bosons in decays of  charged $K$ mesons and neutral $K^0_L$, $K^0_S$ mesons. In the case of  decays of $K^0_L$, $K^0_S$ mesons, we took into account that these particles have certain CP parity and the CS boson is a CP even particle. 
As a result, we found that reaction $K^0_L \rightarrow \pi^0 + X$ has the greatest branching  among the reactions of the CS boson production in decays of kaons. The branching of the reaction $K^0_S \rightarrow \pi^0 + X$ is suppressed in $\sim 10^3$ times compared to the reaction $K^0_L \rightarrow \pi^0 + X$, see  table \ref{tab:BR} and figure~\ref{fig:KBX}(a).\newpage

We consider dominant reactions  of the CS bosons production in two-body  decays of  $B$ meson into  
pseudoscalar mesons ($K$ and  $\pi$ mesons); scalar mesons $K^{0*}(700)$ and $K^{0*}(1430)$; vector mesons $K^*(892)$, $K^*(1410)$, $K^*(1680)$; pseudovector mesons $K_1(1270)$ and $K_1(1400)$;  tensor final meson state $K_2(1430)$. Results for the branching of the corresponding reactions  are presented in table \ref{tab:BR} for a very small mass of the CS boson.  Dependence of the branching  of the CS boson production  on its mass is presented in figure~\ref{fig:KBX}(b),  where, for clarity, we summed up the contributions over the (pseudo)scalar, (pseudo)vector and tensor channels. As one can see, channels of the CS boson production with the greatest branching are decays of $B$ mesons into  pseudovector,  vector and scalar  mesons for the CS boson mass up to $m_X\simeq 3 \rm{\,GeV}$, into  pseudovector and  vector mesons for $3 \rm{\,GeV}\lesssim m_X\lesssim 4 \rm{\,GeV} $, vector and pseudoscalar mesons for $4 \rm{\,GeV}\lesssim m_X\lesssim 4.37 \rm{\,GeV} $, and pseudoscalar mesons for $4.37 \rm{\,GeV}\lesssim m_X\lesssim 4.79 \rm{\,GeV} $. The greatest branching of the CS boson production up to $m_X\simeq 3.8$ GeV is that of  channel $B^{\pm} \to X K^{\pm}_{1}(1270)$, but other channels are also important and can not be neglected.

The results of our research will be helpful for the  construction of the sensitivity region and
the search for the GeV-scale long-lived CS boson at intensity frontier experiments such as MATHUSLA \cite{Curtin:2018mvb}, FACET \cite{Cerci:2021nlb}, FASER \cite{FASER:2018ceo,FASER:2018eoc}, NA62 \cite{Mermod:2017ceo,NA62:2017qcd,Drewes:2018gkc}, DUNE \cite{DUNE:2015lol,DUNE:2020fgq}, etc. Especially considering that the CS boson was already included in the SHiP experiment proposal \cite{Alekhin:2015byh}.

\section*{Acknowledgments}

The authors are grateful to Alexey Boyarsky and Oleg Ruchayskiy for fruitful discussions and helpful comments. We also wish to express thanks to Oleg Barabash for his helpful consultations. We  are grateful to Oleksandr Khasai and Ivan Hrynchak 
for carefully checking the results obtained.

\newpage

\appendix

\section{Computation of the loop diagram for CS boson interaction  with different down-type quarks}\label{appA}

\begin{figure}[h]
\centering
\includegraphics[width=0.5\textwidth]{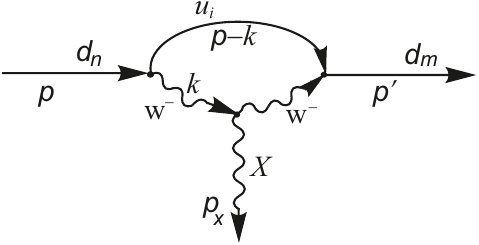} \caption{Interaction of the CS boson with two down-type quarks via loop diagram.}
\label{fig:Xbs}
\end{figure}

The amplitude of a heavy down-type quark ($d_n$) decay into a light down-type quark $(d_m)$  and the CS boson $X_\mu$ in the unitary gauge can be presented as
\begin{equation}\label{Mfi}
    M_{fi}=\frac{g^2}{2}\sum_{i=u,c,t} V_{d_m i}^+V_{id_n}\,\overline{d_m}(p^\prime) \hat P_R  I^{\mu}_{(i)} \hat P_L d_n(p) \epsilon^{*\lambda_X}_{\mu},
\end{equation}
where $\hat P_{R(L)}$ -- projection operators on the right(left)-handed chirality states and in the unitary gauge $I^{ \mu}_{(i)}$ has form
\begin{multline}\label{I}
   I^{ \mu}_{(i)}=\int\frac{{\rm d}^4k}{(2\pi)^4}\gamma^{\alpha}G(p-k)D_{\alpha{ {\rho}}}(k-p_X)\left[c_{\omega}(k-p_X)_{ {\lambda}}+c^{\ast}_{\omega}
   k_{ {\lambda}}\right]D_{ {\nu}\beta}(k)\gamma^{\beta}\epsilon^{{ {\mu}}{ {\nu}}{ {\lambda}}{ {\rho}}}=\\
  \hat A(k) \gamma^{\alpha}\left[m_i+(\not{\!p}-\not{\!k})\right]\left[g_{\alpha{ {\rho}}}-\frac{(k-p_X)_\alpha (k-p_X)_{ {\rho}}}{M_W^2}\right]\left[c_{\omega}(k-p_X)_{ {\lambda}}+c^{\ast}_{\omega}k_{ {\lambda}}\right] \times \\ 
   \left[g_{\beta{ {\nu}}}-\frac{k_\beta k_{ {\nu}}}{M_W^2}\right]\gamma^{\beta}\epsilon^{{ {\mu}}{ {\nu}}{ {\lambda}}{ {\rho}}},
\end{multline}
where $m_i$ is the mass of the $u_i$ quark and integral operator $\hat A$ is defined as
\begin{equation}\label{F}
\hat A(k)=\!\!\int\!\!\frac{{\rm d}^4k}{(2\pi)^4F(k)},\,\,      F(k)=\left[m^2_i-(p-k)^2\right]\left[M_W^2-(k-p')^2\right]\left[M_W^2-k^2\right].
\end{equation}

Relation \eqref{I} can be further simplified with the help of the following identities: 
\begin{equation}\label{p3}
\hat P_R(a+b\gamma_{i})\hat P_L=b \hat P_R \gamma_{i}\hat P_L,\quad  \hat P_R \gamma_{i}\gamma_{j}\hat P_L= 0.
\end{equation}

We use the technique of $\alpha$ (Schwinger) representation, see e.g. \cite{Bogolyubov:1983gp},  namely relations
\begin{equation}\label{Shwinger}
\frac1{m^2-k^2-{\rm i}\varepsilon}={\rm i}\int\limits_0^\infty {\rm d} \alpha\,
{\rm e}^{{\rm i}\alpha(k^2-m^2+{\rm i}\varepsilon)},\quad \varepsilon\rightarrow0,
\end{equation}
\begin{align}
&\int\limits_{-\infty}^\infty {\rm d}^4k\,{\rm e}^{{\rm i}(A k^2+2B k)}=\frac
{\pi^2}{{\rm i}}\cdot\frac1{A^2}\,{\rm e}^{-{\rm i}\frac{B^2}{A}},\label{l21.3a}\\
&\int\limits_{-\infty}^\infty {\rm d}^4k\,{\rm e}^{{\rm i}(A k^2+2B k)}k^\nu
=\frac{\pi^2}{{\rm i}}\cdot\frac1{A^2}\,
{\rm e}^{-{\rm i}\frac{B^2}{A}}\left[-\frac{B^\nu}{A}\right],\label{l21.3b}\\
&\int\limits_{-\infty}^\infty {\rm d}^4k\, {\rm e}^{{\rm i}(A k^2+2B k)}k^\nu
k^\mu=\frac{\pi^2}{{\rm i}}\cdot\frac1{A^2}\,
{\rm e}^{-{\rm i}\frac{B^2}{A}}\left[\frac{2B^\nu B^\mu+{\rm i}A g^{\mu\nu}}{2A^2}\right],\label{l21.3c}\\
& \int\limits_{-\infty}^\infty {\rm d}^4k\, {\rm e}^{{\rm i}(A k^2+2B k)}k^\nu k^\mu
k^\lambda=\frac{\pi^2}{{\rm i}}\cdot\frac1{A^2}\,
{\rm e}^{-{\rm i}\frac{B^2}{A}}\times\nonumber\\
&\hspace{4em} \times\left[-\frac{4B^\mu B^\nu B^\lambda+2{\rm i}A
\left[g^{\mu\nu}B^\lambda+g^{\mu\lambda}B^\nu+g^{\nu\lambda}B^\mu
\right]}{4A^3}\right].\label{l21.3e}
\end{align}
With the help of this technique, we can get the following relations:
\begin{align}
 &  K^{(0)}=\hat A \cdot 1= \frac{{\rm i}\pi^2}{(2\pi)^4}\int_0^{1}\!\!{\rm d} x\!\!\int_0^{1-x}\!\!\!\!{\rm d} y\frac{1}{D(m_i)},\label{G0}\\
&   K_{\alpha}^{(1)}=\hat A k_\alpha =\frac{{\rm i} \pi^2}{(2\pi)^4}\int_0^{1}\!\!{\rm d} x\!\!\int_0^{1-x}\!\!\!\!{\rm d} y\frac{(xp+yp_X)_\alpha}{D(m_i)}, \label{G1}\\
&   K^{(2)}_{\alpha\beta}=\hat A k_\alpha k_\beta
    = \frac{{\rm i} \pi^2}{(2\pi)^4}\int_0^{1}\!\!{\rm d} x\!\!\int_0^{1-x}\!\!\!\!{\rm d} y\left[ \frac{(xp+yp_X)_\alpha (xp+yp_X)_\beta}{D(m_i)}-\frac{g_{\alpha\beta}}{2}\ln{\frac{\Lambda^{2}x}{D(m_i)}}\right],\label{G2}
\end{align}
where $\Lambda$ is some constant with dimension of mass (it should be put to infinity in the end of the computation, $\Lambda\rightarrow \infty$) and
function $D(m_i)$ is defined by \eqref{D1m}.

If we introduce  integral operators $\hat \Lambda_{0(1)}$  \eqref{Lambda0m},  \eqref{Lambda1m} and  notation $\mathcal{P} = xp+yp_X$,
we can rewrite relation  \eqref{G0} -- \eqref{G2} as
\begin{equation}
     K^{(0)}=  \hat \Lambda_0,\quad K^{(1)}_{i}=  \hat \Lambda_0\mathcal{P}_i,\quad K^{(2)}_{i\bar{\lambda}}=  \hat \Lambda_0\mathcal{P}_{\bar{\lambda}}\mathcal{P}_i +\frac{{\rm i}}2\,g_{\bar{\lambda}i}\,\hat \Lambda_1.
\end{equation}
It will be useful also to give the relation
\begin{equation}\label{G3ok}
    K^{(3)}_{i}=\hat A k^2 k_i=\hat \Lambda_0  \mathcal{P}^2 \mathcal{P}_i+3{\rm i} \hat\Lambda_1 \mathcal{P}_i.
\end{equation}

Using the relations obtained above, we can get\vspace{-0.5em}
\begin{multline}\label{unit}
    I^{ \mu}_{(i)}=
   \hat\Lambda_0 \left\{ \vphantom{\frac12}
   \gamma_{ \rho} (\not{\! \mathcal{P}}-\not{\!p})\gamma_{ \nu}\left\{\Theta_{W1}(p_X-2\mathcal{P})_{ \lambda}+{\rm i}\Theta_{W2} p_{X, \lambda} \right\}+\right. \\ +
 \left.  \frac{p_{X, \lambda}}{M_W^2}
      \mathcal{P}_{ \nu}\gamma_{ \rho}\left[c_w^* \left\{2(p\mathcal{P})+\not{\! p_X}\!\!\not{\! \mathcal{P}}-\not{\!p_X}\!\!\not{\! p}\right\}-2\Theta_{W1}  \mathcal{P}^2
      +2{\rm i}\Theta_{W2}  \not{\! p}\not{\! \mathcal{P}}
      \right] \right\} \epsilon^{{ {\mu}}{ {\nu}}{ {\lambda}}{ {\rho}}} +  \\
      + \hat\Lambda_1\left\{ -{\rm i}\Theta_{W1}\gamma_{ {\rho}}\gamma_{ \lambda}\gamma_{ {\nu}}+ \frac{p_{X, \lambda}}{M_W^2} \left[
      {\rm i} \frac{c_w^*}2  \gamma_{ \rho}\not{\!p}_X\gamma_{ \nu} -6{\rm i}\Theta_{W1} \gamma_{ \rho}\mathcal{P}_{ \nu}-\Theta_{W2} \gamma_{ \rho}\not{\!p}\gamma_{ \nu}\right] \right\}
      \epsilon^{{ {\mu}}{ {\nu}}{ {\lambda}}{ {\rho}}}.
\end{multline}

The divergent part of this relation is hidden in the integral operator $\hat\Lambda_1$, see \eqref{Lambda1m}, but we can distinguish in this operator singular and finite parts \eqref{Lambda1regm} and replace 
operator $\hat\Lambda_1$ by finite operator  $\hat\Lambda_1^{fin}$ \eqref{Lambda1finm}, see section~\ref{sec:i}.

With help of the unitary condition for the CKM matrix, it is not difficult to  show that the amplitude of the process does not  depend on the value of  $\mu$. Indeed, for reaction $b\rightarrow s +X$ we have $  V_{su}^+V_{ub} = -( V_{st}^+V_{tb} +  V_{ct}^+V_{cb})  $ and
\begin{multline}
   \sum_{i=u,c,t} V_{si}^+V_{ib} \hat \Lambda_1^{fin} F(x,y, \{p\})\sim 
     \int_0^{1}\!\!{\rm d} x\!\!\int_0^{1-x}\!\!\!\!{\rm d} y \sum_{i=u,c,t} V_{si}^+V_{ib} \ln\frac{D(m_i)}{\mu^2} F(x,y, \{p\}) =\\=
     \int_0^{1}\!\!{\rm d} x\!\!\int_0^{1-x}\!\!\!\!{\rm d} y 
    \left[V_{st}^+V_{tb} \left(\ln\frac{D(m_t)}{\mu^2} - \ln\frac{D(m_u)}{\mu^2}\right)+\right. \\ \left.  + V_{sc}^+V_{cb} \left( \ln\frac{D(m_c)}{\mu^2}-\ln\frac{D(m_u)}{\mu^2}\right) \right]F(x,y,\{p\})=\\
    \int_0^{1}\!\!{\rm d} x\!\!\int_0^{1-x}\!\!\!\!{\rm d} y 
    \left[V_{st}^+V_{tb} \ln\frac{D(m_t)}{D(m_u)}   + V_{sc}^+V_{cb}  \ln\frac{D(m_c)}{D(m_u)} \right]F(x,y,\{p\}). 
\end{multline}
In the following, we will put   $\mu=M_W$  for the convenience of computations.
 
Instead of momentum $p_X$, it is better to use the momentums of quarks $p_X \rightarrow p-p^\prime$. It will allow us to use
relations $\bar d_m(p^\prime)\!\! \not{\! p^\prime} =m^\prime \bar d_m(p^\prime)$ and $\not{\!p}d_n(p) =m d_n (p)$ and, in particular, to get
\begin{align}
& \bar d_m(p^\prime)  \hat P_R \gamma_{\bar\rho} \not{\!p}
\gamma_{\bar\nu} \hat P_L d_n(p)= \bar d_m(p^\prime)  \hat P_R [ 2\gamma_{\bar\rho}\, p_{\bar\nu}  ]\hat P_L d_n(p) - \bar d_m(p^\prime) \hat P_R [m_b \gamma_{\bar\rho} \gamma_{\bar\nu}]\hat  P_R d_n(p),\nonumber \\
&  \bar d_m(p^\prime)  \hat P_R \gamma_{\bar\rho} \not{\!p^{\,\prime}}
\gamma_{\bar\nu} \hat P_L d_n(p)= \bar d_m(p^\prime)  \hat P_R [ 2p^\prime_{\bar\rho}\,\gamma_{\bar\nu}   ]\hat P_L d_n(p) - \bar d_m(p^\prime) \hat P_L [m_s \gamma_{\bar\rho} \gamma_{\bar\nu}]\hat  P_L d_n(p),\nonumber \\
& \bar d_m(p^\prime)  \hat P_R \gamma_{\bar\rho} \not{\!p^{\,\prime}}
\not{\! p} \hat P_L d_n(p)= \bar d_m(p^\prime)  \hat P_R [ 2m_b\, p^\prime_{\bar\rho}   ]\hat P_R d_n(p) - \bar d_m(p^\prime) \hat P_L [m_s m_b \gamma_{\bar\rho} ]\hat  P_R d_n(p),\nonumber
\end{align}
where $m$ is the mass of the initial down-type quark $(d_n)$  and $m^\prime$ is the mass of the final down-type quark $(d_m)$.
After this substitution, we finally get the relation $Y^\mu_{(i)} =\hat P_R I^{\mu}_{(i)} \hat P_L$ \eqref{unitrep1m}.

\section{Coefficients in the amplitude of the CS boson interaction with different down-type quarks}\label{appB}

Let us  perform summation over virtual quarks and integration over the variables $x$ and $y$ in \eqref{Mfim}, \eqref{unitrep1m}.
For simplicity of notation we consider only  decay of the $b $ quark into $s$ quark and the CS boson.
First, it is more convenient  to compute  detached terms that are included in \eqref{Mfim} such as 
  \begin{equation}\label{Lkkm}
 \begin{array}{ll}
  \displaystyle A_0 = \sum_{i=u,c,t} V_{si}^+V_{ib}\, \hat \Lambda_0, & \quad \displaystyle A_{0f} = \sum_{i=u,c,t} V_{si}^+V_{ib}\, \hat \Lambda_0  f,  \\
 \displaystyle A_1 = \sum_{i=u,c,t} V_{si}^+V_{ib} \, \hat \Lambda^{fin}_1, & \quad \displaystyle A_{1f} = \sum_{i=u,c,t} V_{si}^+V_{ib} \, \hat \Lambda^{fin}_1 f.
\end{array}
  \end{equation}
To do it analytically, we will make an estimation, taking into account that $M_W^2
\gg m^2,\, m^{\prime\, 2},\, M_X^2$, and  simplify relation \eqref{D1m} to
the form:
  \begin{equation}\label{Lkkm1}
D(m_i)=\left\{\begin{array}{ll}
    D_{uc}= (1-x)M_W^2 & \quad \mbox{for $i=$ u, c quarks},  \\
    D_t= (1+(t_w-1) x) M_W^2 & \quad \mbox{for $i=$ t quark,}
\end{array}
\right.
  \end{equation}
where $t_w=m_t^2/M_W^2$.
Then one can easily obtain quite simple relations e.g.
\begin{align}
 & A_0 = \frac{{\rm i}\pi^2}{(2\pi)^4}\,\frac1{M_W^2} \left\{  V_{su}^+V_{ub}+V_{sc}^+V_{cb}
   +V_{st}^+V_{tb} \frac{1 - t_w + t_w \ln t_w}{(t_w-1)^2}\right\};\\
 & A_{0x} = \frac{{\rm i}\pi^2}{(2\pi)^4}\,\frac1{2M_W^2} \left\{  V_{su}^+V_{ub}+V_{sc}^+V_{cb}
   +V_{st}^+V_{tb} \frac{t_w^2-1 - 2 t_w \ln t_w}{(t_w-1)^3} \right\};\\
& A_{1} = -\frac14\,\frac{\pi^2}{(2\pi)^4} \left\{ V_{su}^+V_{ub} + V_{sc}^+V_{cb}
   - V_{st}^+V_{tb} \frac{2 t_w^2 \ln t_w -1 + 4 t_w - 3 t_w^2}{(t_w-1)^2} \right\}.
\end{align}

From the unitarity condition $(V^+V)_{sb}=0$, one can conclude
$V_{su}^+V_{ub} + V_{sc}^+V_{cb}=-V_{st}^+V_{tb}$ and write
\begin{align}
 & A_0 = -\frac{{\rm i}\pi^2}{(2\pi)^4}\,\frac{V_{st}^+V_{tb}}{M_W^2}\, t_w\, \frac{ t_w-1 - \ln t_w}{(t_w-1)^2} ;\\
 & A_{0x} = - \frac{{\rm i}\pi^2}{(2\pi)^4}\,\frac{V_{st}^+V_{tb}}{M_W^2}\, t_w \,  \frac{ 3 + (t_w-4) t_w + 2 \ln t_w}{2(t_w-1)^3};\\
 & A_{0y} = -\frac{{\rm i}\pi^2}{(2\pi)^4}\,\frac{V_{st}^+V_{tb}}{M_W^2} \, t_w \,  
        \frac{t_w^2 -1- 2 t_w \ln t_w}{4( t_w-1)^3};\\
&  A_{0x^2} = -\frac{{\rm i}\pi^2}{(2\pi)^4}\,\frac{V_{st}^+V_{tb}}{M_W^2} \, t_w \,  
         \frac{2 t_w^3 - 9 t_w^2 + 18 t_w -11  - 6 \ln t_w}{6 (t_w-1)^4};\\
&  A_{0y^2} = -\frac{i\pi^2}{(2\pi)^4}\,\frac{V_{st}^+V_{tb}}{M_W^2} \, t_w \,  
         \frac{1 - 6 t_w + 3 t_w^2 + 2 t_w^3 - 6 t_w^2 \ln t_w}{18 ( t_w-1)^4};\\
&  A_{0xy} = -\frac{{\rm i}\pi^2}{(2\pi)^4}\,\frac{V_{st}^+V_{tb}}{M_W^2} \, t_w \,  
         \frac{2 + 3 t_w - 6 t_w^2 + t_w^3 + 6 t_w \ln t_w }{12 ( t_w-1)^4};\\
&  A_{0x^3} = -\frac{{\rm i}\pi^2}{(2\pi)^4}\,\frac{V_{st}^+V_{tb}}{M_W^2} \, t_w \,  
         \frac{25 - 48 t_w + 36 t_w^2 - 16 t_w^3 + 3 t_w^4 + 12 \ln t_w}{12(t_w-1)^5};
\end{align}        
\begin{align}      
&  A_{0x^2y} = -\frac{{\rm i}\pi^2}{(2\pi)^4}\,\frac{V_{st}^+V_{tb}}{M_W^2} \, t_w \,  
        \frac{-3 - 10 t_w + 18 t_w^2 - 6 t_w^3 + t_w^4 - 12 t_w \ln t_w}{24(t_w-1)^5};\\
&  A_{0y^2x} = -\frac{{\rm i}\pi^2}{(2\pi)^4}\,\frac{V_{st}^+V_{tb}}{M_W^2} \, t_w \,  
        \frac{-1 + 8 t_w - 8 t_w^3 + t_w^4 + 12 t_w^2 \ln t_w}{36(t_w-1)^5};\\
 & A_{1} = \frac{\pi^2}{(2\pi)^4}\,V_{st}^+V_{tb} \, t_w \,   \frac{1 - t_w + t_w \ln t_w}{2(t_w-1)^2 };\\
 & A_{1x} = \frac{\pi^2}{(2\pi)^4}\,V_{st}^+V_{tb}\,  t_w\, \frac{2 (t_w-1) + ( t_w-3) t_w \ln t_w}{6(t_w-1)^3}.
\end{align}
Denoting 
\begin{equation}
   A_{0\alpha}= -{\rm i}\frac{\pi^2}{(2\pi)^4}V_{st}^+V_{tb}\, t_w \, \frac{a_{0\alpha}}{{M_W^2}}, \quad 
   A_{1\alpha}= \frac{\pi^2}{(2\pi)^4}\,V_{st}^+V_{tb}\,  t_w\, a_{1\alpha},
\end{equation}
where the subscript $\alpha$ is some combination of $x$ and $y$ (or empty index), we get values for dimensionless coefficients $a_{0\alpha}$, $a_{1\alpha}$, see table \ref{tab:coefai}. As one can see, these coefficients do not depend on the down-type quarks which were at the beginning and at the end of the reaction.

\begin{table}[t]
\centering
\resizebox{\textwidth}{!}{%
\begin{tabular}{|c|c|c|c|c|c|c|c|c|c|c|}
\hline
$a_{0}$ &  $a_{0x}$ & $a_{0y}$ &  $a_{0x^2}$ & $a_{0xy}$ &  $a_{0y^2}$
& $a_{0x^3}$ &  $a_{0x^2y}$ &  $a_{0y^2x}$ &  $a_{1}$ &  $a_{1x}$ \\\hline
 0.16  &  0.094 & 0.033 & 0.066 & 0.014 & 0.012 & 0.051 & 0.0076 & 0.0042 & 0.13 & 0.066 \\
\hline
\end{tabular}}
\caption{Numerical values of the coefficients $a_{0\alpha}$ and $a_{1\alpha}$.}\label{tab:coefai}
\end{table}\vspace{-1em}

Now we can write relations for coefficients in \eqref{Mfi2m}:
\begin{align}
 & a_L^d = - {\rm i}\,a_1 \Theta_{W1},  \label{dL}\\
 &  c_L^d= {\rm i} \left(a_{0y} - 2  a_{0y^2} -\frac{a_1}{2}\right) \Theta_{W1} - \left(a_{0y}+\frac{a_1}{2}\right) \Theta_{W2}, \label{bL}\\
 &  d_L^d= {\rm i} \left(2  a_{0xy} - a_{0y} + 2  a_{0y^2} +\frac{a_1}{2}\right) \Theta_{W1} + \left(a_{0y}+\frac{a_1}{2}\right) \Theta_{W2},\label{cL}\\
& b_R^d= - {\rm i}\,   (a_{0x}\! -\! a_{0x^2})  \Theta_{W1} + (a_{0x^2}\! +\! 2 a_{0xy}\!-\!a_{0x})  \Theta_{W2},\label{aR} \\
    &  c_R^d = {\rm i} \left( a_0 - a_{0x} + 2  a_{0xy} - 3 a_{0y} + 2  a_{0y^2} + \frac{a_1}2   \right) \Theta_{W1} \nonumber \\
&\hspace{15em}   - \left(a_0 - a_{0x} - a_{0y} + \frac{a_1}2\right) \Theta_{W2},\label{bR}\\
& d_R^d= {\rm i} \left(- a_0 + 3  a_{0x} - 2  a_{0x^2} - 4  a_{0xy} + 3  a_{0y} - 2  a_{0y^2} - \frac{a_1}2   \right)
  \Theta_{W1} \nonumber \\
&\hspace{15em} + \left(a_0 - a_{0x} - a_{0y} + \frac{a_1}2\right) \Theta_{W2}.\label{cR}
\end{align}

We also write separately the most cumbersome coefficient\vspace{-0.5em}
\begin{multline}
     b_L^d= {\rm i} \left[\vphantom{\frac12} -8 a_{0xy} -8 a_{0y^2} +2 a_{0x} +8 a_{0y} -2 a_0 -2 a_1 +6 a_{1x} + \frac{m_s^2}{M_W^2} [- a_{0xy} +2 a_{0x^2y}]+\right. \\   \left. +
     \frac{m_b^2}{M_W^2} [3 a_{0x^2} + a_{0xy} - a_{0x} -2 a_{0x^3} -2 a_{0x^2y}]   + \frac{M_X^2}{M_W^2} [2 a_{0xy} -2 a_{0x^2y} -2 a_{0xy^2}] \right]  \Theta_{W1}+\\+
     \left[ -2 a_{0x} -4 a_{0y} +2 a_0 + \frac{m_b^2}{M_W^2} [a_{0x^2} + a_{0xy} - a_{0x}] + \frac{m_s^2}{M_W^2} a_{0xy} \right]\Theta_{W2}.
 \end{multline}

The numerical values of these coefficients are presented  in table \ref{tab:coefupdownm}.\newpage

\section{Decays of mesons with the CS boson production}

\subsection{Decay into   mesons of the same parity}\label{app:sameparity}

Let us consider here the production of the CS boson in decays of $B$ mesons  into $K$, $\pi$ mesons  and decays of $K$ mesons into $\pi$ mesons.

In the case of a
decay of one meson into another meson with
the same parity, we have the following relation for the averaging the quark current over the meson states:
\begin{multline}\label{psinps}
   2 \langle h'(p')|\bar{Q}_{i}\gamma^{\mu}\hat P_L Q_{j}|h(p)\rangle = \langle h'(p')|\bar{Q}_{i}\gamma^{\mu}Q_{j}|h(p)\rangle=
    \\ =\left[(p+p')^{\mu} - \frac{m_{P}^{2} - m_{P'}^{2}}{q^{2}}q^{\mu}\right]f^{hh'}_{+}(q^{2})+\frac{m_{P}^{2} - m_{P'}^{2}}{q^{2}}q^{\mu}f^{hh'}_{0}(q^{2}),
\end{multline}
where $q$ is the momentum transfer to $h^\prime$ meson, namely $q = p - p'$, $q^2=M_X^2$. 

Using relations   \eqref{amplitude}, \eqref{epspp}  one can get
the amplitude of the an $h$ meson decay (pseudomeson, $P$) into an $h'$ meson (with the same parity) and the CS boson:
\begin{equation}
    M_{P\rightarrow P'X}=
     \Theta_{1W}  C_{mn}|\vec k| \frac{M_h}{M_X}  f_+^{hh'}(M_X^2).
\end{equation}
Thus, we will be interested only in the $f_+^{hh'}$ form-factor. It should be noted that in the case of  $K^0_S$, $K^0_L$ mesons with a certain CP parity, we have to use the real or the imaginary part of the  coefficient $C_{ds}$, see 
\eqref{amplitudeKS}, \eqref{amplitudeKL}.

We take form-factors $f_+^{hh'}(q^{2})$ for the decays of $B$ mesons  from \cite{Ball:2004ye}, where they were given with the help of  pole parametrizations:
\begin{align}
& f^{hh'}_{+}(q^{2})=\frac{r_1}{1-q^{2}/m_1^{\,2}}+\frac{r_2}{1-q^{2}/m_{fit}^{\,2} }\label{eq:pseudoscalar-form-factor1},\\
& f^{hh'}_{+}(q^{2})=\frac{r_1}{1-q^{2}/m_1^{\,2}}+\frac{r_2}{\left(1-q^{2}/m_1^{\,2}\right)^2 }.\label{eq:pseudoscalar-form-factor2}
\end{align}
The last relation is used for the case when $m_{fit}$ gets too close to $m_1$. 

Form-factors for the decay of $K$ meson are discussed in detail in \cite{Zyla:2020zbs}. 

We use form-factors $f_+^{hh'}(q^{2})$ for the decays of $K^\pm$ mesons in the form of a
quadratic expansion:
\begin{equation}\label{eq:pseudoscalar-form-factor01}
   f^{hh'}_{+}(q^{2})=f^{hh'}_{+}(0)\left(1+\lambda'_+\frac{q^2}{m_{\text{fit}}^{\,2}}+\frac{\lambda''_+}2\left(\frac{q^2}{m_{\text{fit}}^{\,2}}\right)^2\right).
\end{equation}
We take  values of the parameters  $\lambda'_+$ and $\lambda''_+$ from \cite{Zyla:2020zbs}, which are the averaged over the values given in \cite{Yushchenko:2004zs,OKA:2017lbd, NA482:2018rgv}.

For the decays of $K_L^0$ mesons we take form-factors  $f^{hh'}_{+}(q^{2})$ in the form of the 
quadratic expansion \eqref{eq:pseudoscalar-form-factor01} too. We take  values of the parameters  $\lambda'_+$ and $\lambda''_+$ from \cite{Zyla:2020zbs}, which are the averaged  over the values given in \cite{NA48:2004jcz,Alexopoulos:2004sy,KLOE:2006kms}, assuming $\mu-e$  universality.

The values of the parameters in \eqref{eq:pseudoscalar-form-factor1} -- \eqref{eq:pseudoscalar-form-factor01}   are summarized in table~\ref{tab:pseudoscalar-form-factor-parameters}.

For the decays of $K_S^0$ mesons we take form-factors  $f^{hh'}_{+}(q^{2})$ in the form of 
a linear  expansion:
\begin{equation}\label{eq:pseudoscalar-form-factor01linear}
 f^{hh'}_{+}(q^{2})=f^{hh'}_{+}(0)\left(1+\lambda_+\frac{q^2}{m_{\text{fit}}^{\,2}}\right),
\end{equation}
where values of the parameters $f(0)=0.96$ and  $\lambda_+=3.39\cdot 10^{-2}$ were taken from  \cite{KLOE:2006vvm}.

\begin{table}[t]
\begin{center}
$
\begin{array}{|c|rrcr|}
\hline
& r_1 & r_2 & (m_1)^2 & m_{\rm fit}^2\\\hline
f_+^{B\pi} &0.744 & -0.486 & (m_{1}^\pi)^2 & 40.73\\
f_+^{BK} & 0.162 & 0.173 & (m_1^K)^2 & -\\
\hline
\end{array}
$\qquad
$
\begin{array}{|c|cccc|}
\hline
& f(0) & \lambda'_+ & \lambda''_+ & m_{\rm fit}^2\\\hline
f_+^{K^\pm \pi} &0.96 &  2.59\cdot10^{-2}  & 1.86\cdot10^{-3} &m_{\pi^+}^2\\
f_+^{K^0_L \pi} &0.96 &  2.4\cdot10^{-2}  & 2\cdot10^{-3} &m_{\pi^+}^2\\
\hline
\end{array}
$
\end{center}
\caption{Fit parameters for Eqs.~\eqref{eq:pseudoscalar-form-factor1} -- \eqref{eq:pseudoscalar-form-factor01}.
Here $m_1$ is the meson mass in the corresponding channel:
$m_1^{\pi} = m_{B^*} = 5.32$ GeV and $m_1^K = m_{B^*_s} = 5.41$ GeV.}\label{tab:pseudoscalar-form-factor-parameters}
\end{table}

\subsection{Decays into  mesons of another parity}\label{app:anotherparity}

Let us consider here the production of the CS boson in the decays of $B$ mesons into $K^{0*}(700)$, $K^{0*}(1430)$ mesons.

In the case of a
decay of one meson into another meson with
different  parity we get
\begin{equation}\label{psins}
    2\langle h^\prime(p^\prime)|\bar{Q}_{i}\gamma^{\mu}\hat P_L Q_{j}|h(p)\rangle = - \left[(p+p')^{\mu} -q^{\mu}\right]f^{hh'}_{+}(q^{2})=-2p'^{\mu}f^{hh'}_{+}(q^{2}),
\end{equation}
where we used $f_{+}(q^{2}) = -f_{-}(q^{2})$ in \eqref{psinps}, \cite{Sun:2010nv}.

There is an open question whether hypothetical $K_{0}^{*}(700)$ is a state formed by two or four quarks, see, e.g.~\cite{Daldrop:2012sr},  discussions in~\cite{Sun:2010nv,Cheng:2013fba} and references therein. In this paper, we will do the same as we did in \cite{Boiarska:2019jym}, namely, we assume that $K_{0}^{*}(700)$ is a di-quark state and $K_{0}^{*}(1430)$ is its excited state. There are no experimentally observed decays of $B$ meson into $ K_{0}^{*}(700)$, and therefore there is quite a large theoretical uncertainty in the determination of the form-factors (see the discussion in~\cite{Issadykov:2015iba}). We will use~\cite{Sun:2010nv}, where there are results for $B\to K_{0}^{*}(700)$ and $B\to K_{0}^{*}(1430)$, and the results for the latter are in good agreement with the experimental data for $B \to K_{0}^{*}(1430)\eta'$ decay. 

Using relations  \eqref{amplitude}, \eqref{epspp} one can get
the amplitude of an $h$ meson decay (pseudomeson, $P$) into an $h'$ meson (scalar meson, $S$) and the CS boson: 
\begin{equation}
    M_{P\rightarrow SX}=
     -\Theta_{1W}   C_{mn} |\vec k|\frac{M_h}{M_X}  f_+^{hh'}(M_X^2) .
\end{equation}

We take  $f^{BK_{0}^{*}}_{+}(q^{2})$ from~\cite{Sun:2010nv} in the form of a pole-like function:
\begin{equation}
    f^{BK_{0}^{*}}_{+}(q^{2}) = \frac{F^{BK_{0}^{*}}_{0}}{1-a\frac{q^{2}}{m_{B}^{2}}+b\left(\frac{q^{2}}{m_{B}^{2}}\right)^{2}},
    \label{eq:scalar-form-factor}
\end{equation}
where $m_{B} = 5.3\text{ GeV}$ is the mass of the $B^{+}$ meson.
The fit parameters are given in table~\ref{tab:scalar-form-factor-parameters}.

\begin{table}[t]
    \centering
    \begin{tabular}{|c|c|c|c|c|}
        \hline  
        ${S}$ &$F_{0}^{B{S}}$ &    $a$ & $b$   \\
        \hline
         $K_{0}^{*}(700)$ & $0.466$ & $1.501$ & $1.026$ \\
        \hline
        $K_{0}^{*}(1430)$ & $0.181$  & $4.293$  & $6.450$ \\
        \hline
    \end{tabular}
    \caption{Values of the parameters in the form-factor parameterization~\eqref{eq:scalar-form-factor} for $B = B^{+}$, $S = K_{0}^{*0}(700)$, $K_{0}^{*}(1430)$ in regions $q^2<11$ GeV${^2}$ and $q^2<8$ GeV${^2}$ correspondingly. We used figure 3  of \protect\cite{Sun:2010nv} to find interpolation coefficients of \eqref{eq:scalar-form-factor}.}
    \label{tab:scalar-form-factor-parameters}
\end{table}

\subsection{Vector and pseudovector final meson state}

\subsubsection{Vector}
\label{app:vector}

Let us consider here the production of the CS boson in the decays of $B$ mesons into  vector states $B\to K^{*}(892)$, $K^{*}(1410)$, $K^{*}(1680)$. Since the total angular momentum of the B meson is zero, two final vector particles must have zero  total angular momentum.

For the vector final state, $h' = V$, we have~\cite{Ball:2004rg,Ebert:1997mg}
\begin{multline}
    \langle V(p')| \bar{Q}_{i}\gamma^{\mu}\gamma_{5}Q_{j} |h(p)\rangle = (M_h+M_{V})\epsilon_V^{\mu *}(p')A_{1}(q^{2}) - \\ -(\epsilon_V^{*}(p')\cdot q)(p + p')^{\mu}\frac{A_{2}(q^{2})}{M_h+M_{V}}-2M_{V}\frac{\epsilon_V^{*}(p')\cdot q}{q^{2}}q^{\mu}(A_{3}(q^{2})-A_{0}(q^{2})),
    \label{eq:vector-form-factor-axial-part}
\end{multline}
\begin{equation}
     \langle V(p')| \bar{Q}_{i}\gamma^{\mu}Q_{j} |h(p)\rangle= \frac{2V(q^{2})}{M_h+M_{V}}{\rm i}\epsilon^{\mu\nu\rho\sigma}\epsilon^{*}_{V,\nu}(p')p_{ \rho}p'_{\sigma},
     \label{eq:vector-form-factor-vector-part}
\end{equation}
where $\epsilon_V^{\mu}(p')$ is the polarization vector of the vector meson, and $A_{i}, V$ are the form-factors. The form-factor $A_{3}$ is related to $A_{1}$ and $A_{2}$ as
\begin{equation}
    A_{3}(q^{2}) = \frac{M_{h}+M_{V}}{2M_{V}}A_{1}(q^{2})-\frac{M_{h} - M_{V}}{2M_{V}}A_{2}(q^{2}).
    \label{eq:a3-a0}
\end{equation}

The amplitude of an $h$ meson decay into a vector $V$ meson and the CS boson has form\newpage
\begin{multline}\label{amplitudeV}
    M_{h\rightarrow VX}(\lambda_V,\lambda_X)=\Theta_{W1}  g^2   C_{mn} \, \langle V(p',\lambda_V)|\bar d_n\gamma^{\mu} \hat P_L d_m|h(p)\rangle\, \epsilon_{X,\mu}^{*\lambda_X}=\\
    =-\Theta_{W1}  \frac{g^2   C_{mn}}2\left[-\frac{2V(q^{2})}{M_h+M_{V}}{\rm i}\epsilon^{\mu\nu\rho\sigma}\epsilon_{X,\mu}^{\lambda_X*}\epsilon^{\lambda_V*}_{V,\nu}p_{ \rho}p'_{\sigma}+
    (M_{h}+M_V) (\epsilon_V^{\lambda_V *}\cdot \epsilon_X^{\lambda_X *}) A_{1}(q^{2})\right. - \\  \left.-(\epsilon_V^{\lambda_V *}\cdot q)(\epsilon_X^{\lambda_X *}\cdot (p + p'))\frac{A_{2}(q^{2})}{M_{h}+M_{V}}
    -2M_{V}\frac{\epsilon_V^{\lambda_V*}\cdot q}{q^{2}} (\epsilon_X^{\lambda_X *}\cdot q) (A_{3}(q^{2})-A_{0}(q^{2}))
    \right],
\end{multline}
where $\epsilon_V^{\lambda_V *}$ and $\epsilon_X^{\lambda_X *}$ correspond to the polarizations of the vector meson ($V_\mu$) and the CS boson ($X_\mu$).

In the rest frame of the $h$ meson $p=(M_h,0)$ we have relations \eqref{p'pX}.
Let us guide Z-axis along the spatial momentum of the CS boson $\vec k$, then the other vector particle will move in the opposite direction. 
In this case, 4-vectors of the polarizations for the CS boson or the vector particle have form
\begin{equation}\label{polaris1}
    \epsilon_{X(V)}^{(\pm)}=\frac1{\sqrt{2}}\left(0, 1,\mp {\rm i},0\right),\quad
    \epsilon_X^{(0)}=\frac1{M_X}\left(|\vec k|, 0,0,E_X\right),\quad
    \epsilon_V^{(0)}=\frac1{M_V}\left(|\vec k|, 0,0,-E_V\right), 
\end{equation}
where $\epsilon^{(\pm)}$ corresponds to the spin projection $\pm1$ and $\epsilon^{(0)}$ corresponds to  zero spin projection on Z-axis.

Using \eqref{polaris1} one can get \eqref{epspp} and the following relations  for different polarizations of vector particles:
\begin{align}
&     \epsilon_V^{(0)*}\cdot p= |\vec k|\frac{M_h}{M_V}, \,\,\, \epsilon_V^{ (0)*}\cdot p'=0,\nonumber \\
&    \epsilon_V^{(0)*}\cdot \epsilon_X^{(0)*}=\frac{\vec k\,{}^2+E_X E_V}{M_X M_V},\,\,\, \epsilon_V^{(+)*}\cdot \epsilon_X^{(-)*}=\epsilon_V^{(-)*}\cdot \epsilon_X^{(+)*}=-1,\label{prodeps}
\end{align}
where module $|\vec k|$ is defined by \eqref{brdef}.

Consider now the following convolution $\epsilon^{\mu\nu\rho\sigma}\epsilon_{X,\mu}^{*}\epsilon^{*}_{V,\nu}(p')p_{ \rho}p'_{\sigma}$. 
Due to the fact that only the 0 and 3 components of the momentums $p$ and $p'$ are nonzero, only the 1 and 2 components of polarization vectors can be used. These components are zero for the longitudinal polarization of the vector particles, so the contribution from  longitudinal polarizations is absent, but the contribution from transversal opposite polarizations is nonzero:
\begin{equation}\label{levichi}
    \epsilon^{\mu\nu\rho\sigma}\epsilon_{X,\mu}^{(\pm)*}\epsilon_{V,\nu}^{(\mp)*}p_{ \rho}p'_{\sigma}=(p_0 p'_3-p_3 p'_0)\left(\epsilon_{X,1}^{(\pm)*}\epsilon_{V,2}^{(\mp*)}-\epsilon_{X,2}^{(\pm)*}\epsilon_{V,1}^{(\mp)*}\right), 
\end{equation}
 namely
\begin{equation}\label{levivalue}
    \epsilon^{\mu\nu\rho\sigma}\epsilon_{X,\mu}^{(+)*}\epsilon_{V,\nu}^{(-)*}p_{ \rho}p'_{\sigma}=-{\rm i} M_h |\vec k|,\quad  \epsilon^{\mu\nu\rho\sigma}\epsilon_{X,\mu}^{(-)*}\epsilon_{V,\nu}^{(+)*}p_{ \rho}p'_{\sigma}=+{\rm i} M_h |\vec k|.
\end{equation}

So, we  get the following nonzero amplitudes of the reaction:
\begin{align}
 &   M_{h\rightarrow VX}(+,-)=
    -\Theta_{1W}  \frac{  C_{mn}}2\left[\frac{2M_h |\vec k|}{M_h+M_{V}}V(q^{2})-
    (M_{h}+M_V) A_{1}(q^{2})    \right],\\
&    M_{h\rightarrow VX}(-,+)=
    \Theta_{1W}  \frac{   C_{mn}}2\left[\frac{2M_h |\vec k|}{M_h+M_{V}}V(q^{2})+
    (MM_V) A_{1}(q^{2})    \right], \label{amplitudemnpl}\\
&    M_{h\rightarrow VX}(0,0)=
    -\Theta_{1W} \frac{   C_{mn}}2\left[(M_{h}+M_V) \frac{\vec k\,{}^2+E_X E_V}{M_X M_V} A_{1}(q^{2})-2\frac{\vec k\,{}^2M_h^2}{M_V M_X}\frac{A_{2}(q^{2})}{M_{h}+M_{V}}    \right],
\end{align}
and 
\begin{multline}\label{mfivectorsum}
    \sum_{polarizations} \left|M_{h\rightarrow VX}\right|^2=\Theta_{1W}^2 \frac{   |C_{mn}|^2}4\left[ \frac{8M_h^2 \vec k\,{}^2}{(M_h+M_{V})^2}V^2(q^{2})+
    2(M_{h}+M_V)^2 A^2_{1}(q^{2}) +\right.\\  \left. + \left(\!\!(M_{h}+M_V)\frac{\vec k\,{}^2+E_X E_V}{M_X M_V}  A_{1}(q^{2})-\frac{2M_h^2\vec k\,{}^2}{M_V M_X(M_{h}+M_{V})} A_{2}(q^{2})   \!\!\right)^2  \right].
\end{multline}

\begin{table}[t!]
    \centering
    \begin{tabular}{|c|c|c|c|c|c|}
        \hline  
        form-factors & $r_{1}$ & $r_{2}$ & $m_{R}$, GeV & $m_{\text{fit}}, \text{ GeV}$ & $q^2=0$\\
        \hline
        $V(q^2)  $ & $0.923 $ & $-0.511$ & $5.32$ & $\sqrt{49.40}$ & $0.411 $ \\
        $A_1(q^2)$ & $-$ & $0.29$ & $-$ & $\sqrt{40.38}$ & $0.292$ \\
        $A_2(q^2)$ & $-0.084$ & $0.342$ & $-$ & $\sqrt{52.00}$ & $0.259$ \\
        \hline
    \end{tabular}
    \caption{Values of  parameters of the vector form-factors \eqref{eq:vector-form-factor-1} -- \eqref{eq:vector-form-factor-A2} for the decay of a  $B$ meson into $K^{*}(892)$  \protect\cite{Ball:2004rg}.}
    \label{tab:vector-form-factor-parameters1}
\end{table}

For the case of the decay of a  $B$ meson into $ K^{*}(892)$, we follow~\cite{Ball:2004rg} and parametrize the form-factor $V$ and $A_0$ as
\begin{equation}
    F(q^{2}) = \frac{r_{1}}{1-q^{2}/m_{R}^{2}}+\frac{r_{2}}{1-q^{2}/m_{\text{fit}}^{2}},
    \label{eq:vector-form-factor-1}
\end{equation}
form-factor $A_1$ as
\begin{equation}
    F(q^{2}) = \frac{r_{2}}{1-q^{2}/m_{\text{fit}}^{2}},
    \label{eq:vector-form-factor-A1}
\end{equation}
and form-factor $A_2$ as
\begin{equation}
    F(q^{2}) = \frac{r_{1}}{1-q^{2}/m_{\text{fit}}^{2}}+\frac{r_{2}}{\left(1-q^{2}/m_{\text{fit}}^{2}\right)^2}.
    \label{eq:vector-form-factor-A2}
\end{equation}
The values of the corresponding parameters are given in table~\ref{tab:vector-form-factor-parameters1}.

For the case of the decay of a $B$ meson into $ K^{*}(1410), K^{*}(1680)$, we use another parametrizaton for the form-factors~\cite{Hatanaka:2009gb,Lu:2011jm}: 
\begin{align}
&     A_{0}(q^{2}) = \left( 1-\frac{2M_{V}^{2}}{M_h^2 + M_V^2 - q^2}\right)\zeta_{||}(q^{2})+\frac{M_{V}}{M_{h}}\zeta_{\perp}(q^{2}),
     \label{eq:vector-form-factor-2}\\
&   A_1(q^2)
= \frac{2E_{V}}{M_h+M_V} \zeta_{\perp}(q^2) = \frac{M_h^2 + M_V^2 - q^2}{M_h(M_h+M_V)} \zeta_{\perp}(q^2),\\
&   A_2(q^2)
= \left(1+\frac{M_V}{M_h}\right) \left[
\zeta_{\perp}(q^2) - \frac{2M_h M_V}{M_h^2 + M_V^2 - q^2} \zeta_{||} (q^2)
\right],\\
& V(q^2)\label{eq:vector-form-factor-2-2}
= \left(1+\frac{M_V}{M_h}\right) \zeta_{\perp}(q^2),
\end{align}
where 
\begin{equation}
\zeta_{\perp/||}(q^{2}) = \frac{\zeta_{\perp/||}(0)}{1-q^{2}/M_{h}^{2}}.
\end{equation}
The values of the corresponding parameters are given in  table~\ref{tab:vector-form-factor-parameters2}.

\begin{table}[t!]
    \centering
    \begin{tabular}{|c|c|c|}
        \hline  
        $V$ & $\zeta_{\perp}(0)$ & $\zeta_{||}(0)$  \\
        \hline
        $K^{*}(1410)$ & $0.28$ & $0.22$ \\
        $K^{*}(1680)$ & $0.24$ & $0.18$ \\
        \hline
    \end{tabular}
    \caption{Values of  parameters in the vector form-factors~\eqref{eq:vector-form-factor-2} -- \eqref{eq:vector-form-factor-2-2} for the decay of a $B$ meson into $K^{*}(1410), K^{*}(1680)$, \protect\cite{Hatanaka:2009gb,Lu:2011jm}.}
    \label{tab:vector-form-factor-parameters2}
\end{table}

\subsubsection{Pseudo-vector}
\label{app:pseudovector}

For the case of  pseudo-vector mesons, $h' = A$,
one has to interchange the expressions for the vector and axial-vector matrix elements \eqref{eq:vector-form-factor-axial-part},~\eqref{eq:vector-form-factor-vector-part}, see \cite{Bashiry:2009wq,Hatanaka:2008gu}:
\begin{multline}
    \langle A(p')| \bar{Q}_{i}\gamma^{\mu}Q_{j} |h(p)\rangle = (M_h+M_{A})\epsilon_A^{\mu *}(p')V_{1}(q^{2}) - \\ -(\epsilon_A^{*}(p')\cdot q)(p + p')^{\mu}\frac{V_{2}(q^{2})}{M_h+M_{A}}-2M_{A}\frac{\epsilon_A^{*}(p')\cdot q}{q^{2}}q^{\mu}(V_{3}(q^{2})-V_{0}(q^{2})),
    \label{eq:pseudovector-form-factor-axial-part}
\end{multline}
\begin{equation}
     \langle A(p')| \bar{Q}_{i}\gamma^{\mu}\gamma_5 Q_{j} |h(p)\rangle= \frac{2A(q^{2})}{M_h+M_{A}}{\rm i}\epsilon^{\mu\nu\rho\sigma}\epsilon^{*}_{A,\nu}(p')p_{ \rho}p'_{\sigma},
     \label{eq:pseudovector-form-factor-vector-part}
\end{equation}
with the same relation between $V_i$ as for $A_i$ in the case of vector mesons~\eqref{eq:a3-a0}. 

Expression \eqref{mfivectorsum} in this case takes form
\begin{multline}\label{mfipseudovectorsum}
    \sum_{polarizations} \left|M_{h\rightarrow AX}\right|^2=\Theta_{1W}^2  \frac{  |C_{mn}|^2}4\left[ \frac{8M_h^2 \vec k\,{}^2}{(M_h+M_{A})^2}A^2(q^{2})+
    2(M_{h}+M_A)^2 V^2_{1}(q^{2}) +\right.\\  \left. + \left(\!\!(M_{h}+M_A)\frac{\vec k\,{}^2+E_X E_A}{M_X M_A}  V_{1}(q^{2})-\frac{2M_h^2\vec k\,{}^2}{M_A M_X(M_{h}+M_{A})} V_{2}(q^{2})   \!\!\right)^2  \right].
\end{multline}

\begin{table}[t]
\centering
\begin{tabular}{|clll|clll|}
\hline
     $F$
    & $F(0)$
    & $a$
    & $b$
    & $F$
    & $F(0)$
    & $a$
    & $b$
 \\
    \hline
$V_1^{BK_{1A}}$
    & $0.34$
    & $0.635$
    & $0.211$
&$V_1^{BK_{1B}}$
    & $-0.29$
    & $0.729$
    & $0.074$
    \\
$V_2^{BK_{1A}}$
    & $0.41$
    & $1.51$
    & $1.18$
&$V_2^{BK_{1B}}$
    & $-0.17$
    & $0.919$
    & $0.855$
    \\
$A^{BK_{1A}}$
    & $0.45$
    & $1.60$
    & $0.974$
&$A^{BK_{1B}}$
    & $-0.37$
    & $1.72$
    & $0.912$
\\ \hline
\end{tabular}
\caption{Form-factors for $B\to K_{1A},K_{1B}$ transitions  are fitted to the
3-parameter form in \eqref{eq:pseudovector-form-factors-unphys}, see \cite{Hatanaka:2008gu}.} \label{tab:FFinLF}
\end{table}

We will consider decays of $B$ mesons in two lightest pseudo-vector resonances $K_{1}(1270)$, $K_{1}(1400)$, each of which is a mixture of unphysical $K_{1A}$ and $K_{1B}$ states~\cite{Bashiry:2009wq},
\begin{equation}
    \begin{pmatrix} |K_{1}(1270) \rangle \\ |K_{1}(1400) \rangle \end{pmatrix} =  \begin{pmatrix} \sin(\theta_{K_{1}}) & \cos(\theta_{K_{1}}) \\ \cos(\theta_{K_{1}}) & -\sin(\theta_{K_{1}})\end{pmatrix} 
    \begin{pmatrix} |K_{1A} \rangle \\ |K_{1B}\rangle \end{pmatrix}.
\end{equation}
The form-factors $V_{i}^{BK_{1}}$ and $A^{BK_{1}}$ can be related to the appropriate form-factors  of the $K_{1A}$ and $K_{1B}$ states as 
\begin{align}
& A^{BK_{1}(1270)}(q^{2}) =\sin(\theta_{K_{1}})\frac{m_B+m_{K_{1}(1270)}}{m_B+m_{K_{1A}}}A^{K_{1A} }( q^{2})+\cos(\theta_{K_{1}})\frac{m_B+m_{K_{1}(1270)}}{m_B+m_{K_{1B}}} A^{K_{1B} }( q^{2}), \label{eq:pseudovector-form-factors1} \\
& A^{BK_{1}(1400)}(q^{2}) =\cos(\theta_{K_{1}}) \frac{m_B+m_{K_{1}(1400)}}{m_B+m_{K_{1A}}}A^{K_{1A} }( q^{2})-\sin(\theta_{K_{1}}) \frac{m_B+m_{K_{1}(1400)}}{m_B+m_{K_{1B}}} A^{K_{1B} }( q^{2}),
    \label{eq:pseudovector-form-factors2}
\end{align}    
\begin{align}    
& V_2^{BK_{1}(1270)}(q^{2}) =\sin(\theta_{K_{1}})\frac{m_B+m_{K_{1}(1270)}}{m_B+m_{K_{1A}}}V_2^{K_{1A} }( q^{2})+\cos(\theta_{K_{1}})\frac{m_B+m_{K_{1}(1270)}}{m_B+m_{K_{1B}}} V_2^{K_{1B} }( q^{2}), \label{eq:pseudovector-form-factors1V2} \\
& V_2^{BK_{1}(1400)}(q^{2}) =\cos(\theta_{K_{1}}) \frac{m_B+m_{K_{1}(1400)}}{m_B+m_{K_{1A}}}V_2^{K_{1A} }( q^{2})-\sin(\theta_{K_{1}}) \frac{m_B+m_{K_{1}(1400)}}{m_B+m_{K_{1B}}} V_2^{K_{1B} }( q^{2}),
    \label{eq:pseudovector-form-factors2V2}\\
& V_1^{BK_{1}(1270)}(q^{2}) =\sin(\theta_{K_{1}})\frac{m_B+m_{K_{1A}}}{m_B+m_{K_{1}(1270)}} V_1^{K_{1A} }( q^{2}) +\cos(\theta_{K_{1}})\frac{m_B+m_{K_{1B}}}{m_B+m_{K_{1}(1270)}}  V_1^{K_{1B} }( q^{2}), \label{eq:pseudovector-form-factors1V1} \\
& V_1^{BK_{1}(1400)}(q^{2}) =\cos(\theta_{K_{1}}) \frac{m_B+m_{K_{1A}}}{m_B+m_{K_{1}(1400)}} V_1^{K_{1A} }( q^{2})-\sin(\theta_{K_{1}}) \frac{m_B+m_{K_{1B}}}{m_B+m_{K_{1}(1400)}}  V_1^{K_{1B} }( q^{2}),
    \label{eq:pseudovector-form-factors2V1}
\end{align}
where we take $\theta_{K_1}=-34^{\circ}$, $m_{K_{1A}}=1.31\text{ GeV}$,  $m_{K_{1B}}=1.34\text{ GeV}$ and  the momentum dependence of all form-factors
is parameterized as
\begin{equation}
    F(q^{2}) = \frac{F(0)}{1-a\frac{q^{2}}{m_{B}^{2}}+b\left(\frac{q^{2}}{m_{B}^{2}}\right)^{2}}.
    \label{eq:pseudovector-form-factors-unphys}
\end{equation}
The values of all relevant parameters are given in table \ref{tab:FFinLF}.

\subsection{Tensor final meson state}
\label{app:tensor}

Let us consider here the production of the CS boson in the decays of $B$ mesons into  tensor  state $K_{2}^{*}(1430)$.
For the tensor meson, $h' = T$, the expansions of the matrix elements  are similar to~\eqref{eq:vector-form-factor-axial-part},~\eqref{eq:vector-form-factor-vector-part}, see \cite{Cheng:2010yd,Li:2010ra},
    \begin{multline}
    \langle T(p')| \bar{Q}_{i}\gamma^{\mu}\gamma_{5}Q_{j} |h(p)\rangle = (M_{ h}+M_{T})\epsilon^{\mu *,s}_{T}(p')A_{T1}(q^{2}) - \\ -(\epsilon^{*,s}_{T}(p')\cdot q)(p + p')^{\mu}\frac{A_{T2}(q^{2})}{M_{ h}+M_{T}}-2M_{T}\frac{\epsilon^{*,s}_{T}(p')\cdot q}{q^{2}}q^{\mu}(A_{T3}(q^{2})-A_{T0}(q^{2})),
    \label{eq:tensor-form-factor-axial-part}
\end{multline}
\begin{equation}
     \langle T(p')| \bar{Q}_{i}\gamma^{\mu}Q_{j} |h(p)\rangle= \frac{2T(q^{2})}{M_h+M_{T}}{\rm i}\epsilon^{\mu\nu\rho\sigma}\epsilon^{*,s}_{T,\nu}(p')p_{ \rho}p'_{\sigma},
     \label{eq:tensor-form-factor-vector-part}
\end{equation}
with the same relation between $A_{Ti}$ as for $A_i$ in the case of vector mesons~\eqref{eq:a3-a0}. So, the amplitude of the $h$ meson decay into a tensor $T$ meson and the CS boson has form like \eqref{amplitudeV}:
\begin{multline}\label{amplitudeT}
    M_{h\rightarrow TX}(s,\lambda_X)=\Theta_{1W}    C_{mn} \, \langle T(p',s)|\bar d_n\gamma^{\mu} \hat P_L d_m|h(p)\rangle\, \epsilon_{X,\mu}^{*\lambda_X}=\\
    =-\Theta_{1W}  \frac{   C_{mn}}2 \left[-\frac{2T(q^{2})}{M_h+M_{T}}{\rm i}\epsilon^{\mu\nu\rho\sigma}\epsilon_{X,\mu}^{\lambda_X*}\epsilon^{s*}_{T,\nu}p_{ \rho}p'_{\sigma}+
    (M_{h}+M_T) (\epsilon_T^{s *}\cdot \epsilon_X^{\lambda_X *}) A_{1T}(q^{2})\right. - \\  \left.-(\epsilon_T^{s *}\cdot q)(\epsilon_X^{\lambda_X *}\cdot (p + p'))\frac{A_{2T}(q^{2})}{M_{h}+M_{T}}
    -2M_{T}\frac{\epsilon_T^{s*}\cdot q}{q^{2}} (\epsilon_X^{\lambda_X *}\cdot q) (A_{3T}(q^{2})-A_{0T}(q^{2}))
    \right],
\end{multline}
where  $\epsilon_X^{\lambda_X *}$ corresponds to the CS boson polarization, $s = \pm 2, \pm 1, 0$  are the polarization states of the tensor meson and
 $\epsilon^{s}_{T\mu}(p')$ is a vector defined by
\begin{equation}
    \epsilon_{T\mu}^{s}(p') \equiv  \frac{1}{M_{h}}\epsilon_{\mu\nu}^{s}(p')p^{\nu},
\end{equation}
where $\epsilon_{\mu\nu}^{s}$ is the polarization tensor of $T$ meson satisfying conditions $p_{\mu}\epsilon^{\mu\nu,s}(p) = 0$ and $\epsilon^{\mu\nu,s} = \epsilon^{\nu\mu,s}$, $\epsilon^{\mu, \ s}_{\ \mu} = 0$. 
Tensor $\epsilon_{\mu\nu}^{s}$  can be constructed with the help of a spin-1 polarization vectors:
\begin{eqnarray}
 &&\epsilon_{\mu\nu}^{(\pm2)}=
 \epsilon_\mu^{(\pm)}\epsilon_\nu^{(\pm)},\;\;\;\;
 \epsilon_{\mu\nu}^{(\pm1)}=\frac{1}{\sqrt2}
 [\epsilon_{\mu}^{(\pm)}\epsilon_\nu^{(0)}+\epsilon_{\nu}^{(\pm)}\epsilon_\mu^{(0)}],\nonumber\\
 &&\epsilon_{\mu\nu}^{(0)}=\frac{1}{\sqrt6}
 [\epsilon_{\mu}^{(+)}\epsilon_\nu^{(-)}+\epsilon_{\nu}^{(+)}\epsilon_\mu^{(-)}]
 +\sqrt{\frac{2}{3}}\epsilon_{\mu}^{(0)}\epsilon_\nu^{(0)}.
\end{eqnarray}
In the case of the rest frame of an $h$ meson we have  $p=(M_h,0)$ and relations \eqref{p'pX}. We can guide the CS boson to move   along $Z$-axis and the tensor meson to move in the opposite direction. In this case we can choose spin-1 polarization vectors for the tensor meson as
\begin{eqnarray}
\epsilon_\mu^{(0)}&=&\frac{1}{M_{T}}(|\vec
k|,0,0,-E_{T}),\;\;\;
\epsilon_\mu^{(\pm)}=\frac{1}{\sqrt{2}}(0,1,\mp {\rm i},0),
\end{eqnarray}
then we get 
\begin{equation}\label{epsT}
    \epsilon^{\pm 2}_{T\mu} = 0, \quad \epsilon_{T\mu}^{\pm 1} = \frac{|\vec k|}{\sqrt{2}M_T}\epsilon^{(\pm )}_{\mu}, \quad \epsilon^{0}_{T\mu} = \sqrt{\frac{2}{3}} \frac{|\vec k|}{M_{T}} \epsilon_{\mu}^{(0)},
\end{equation}
where module $|\vec k|$ is defined by \eqref{brdef}.

So, the amplitude of the reaction is zero for $s=\pm2$. Using relations \eqref{prodeps}, \eqref{levivalue}, \eqref{epsT}  we get 
\begin{equation}
    M_{h\rightarrow TX}(+,-)=
    -\Theta_{1W} \frac{   C_{mn}}2 \frac{|\vec k|}{\sqrt{2}M_T}\left[\frac{2M_h |\vec k|}{M_h+M_{T}}T(q^{2})-
    (M_{h}+M_T) A_{1T}(q^{2})    \right],
\end{equation}
\begin{equation}\label{amplitudemnplT}
    M_{h\rightarrow TX}(-,+)=
    \Theta_{1W}  \frac{  C_{mn}}2 \frac{|\vec k|}{\sqrt{2}M_T}\left[\frac{2M_h |\vec k|}{M_h+M_{T}}T(q^{2})+
    (M_{h}+M_T) A_{1T}(q^{2})    \right].
\end{equation}
\begin{multline}
    M_{h\rightarrow TX}(0,0)=\\
    -\Theta_{1W}  \frac{ C_{mn}}2 \sqrt{\frac23} \frac{|\vec k|}{M_{T}}\left[(M_{h}+M_T) \frac{\vec k\,{}^2+E_X E_T}{M_X M_T} A_{1T}(q^{2})-2\frac{\vec k\,{}^2M_h^2}{M_T M_X}\frac{A_{2T}(q^{2})}{M_{h}+M_{T}}    \right],
\end{multline}
and 
\begin{multline}\label{mfivectorsumT}
    \sum_{polarizations} \left|M_{h\rightarrow TX}\right|^2=\\=\Theta_{1W}^2 \frac{    |C_{mn}|^2}4 \frac{\vec k\,{}^2}{M_{T}^2}\left[ \frac{1}{2}\left( \frac{8M_h^2 \vec k\,{}^2}{(M_h+M_{T})^2}T^2(q^{2})+
    2(M_{h}+M_T)^2 A^2_{1T}(q^{2})\right) +\right.\\  \left. + \frac23 \left(\!\!(M_{h}+M_T)\frac{\vec k\,{}^2+E_X E_T}{M_X M_T}  A_{1T}(q^{2})-\frac{2M_h^2\vec k\,{}^2}{M_T M_X(M_{h}+M_{T})} A_{2T}(q^{2})   \!\!\right)^2  \right].
\end{multline}

\begin{table}[t]
\begin{center}
\begin{tabular}{| l c c c || c c c c |}
\hline 
$F$
    & $F(0)$
    & $a$
    & $b$
&  $F$
    & $F(0)$
    & $a$
    & $b$
 \\
    \hline
$T^{BK_2^*}$
    & 0.29
    & 2.17
    & $2.22$
& $A_0^{BK_2^*}$
    & $0.23$
    & $1.23$
    & $0.74$ \\
$A_1^{BK_2^*}$
    & $0.22$
    & 1.42
    & $0.50$
& $A_2^{BK_2^*}$
    & 0.21
    & 1.96
    & $1.79$ \\
    \hline
\end{tabular}
\end{center}
\caption{Values of the form-factors' parameters for transition $B\to K_{2}^{*}(1430)$, \protect\cite{Cheng:2010yd}.} \label{tab:FFinCLFQ}
\end{table}

The parametrization of the form-factors $T$ and $A_{1T}$, $A_{2T}$   is taken from \cite{Cheng:2010yd,Li:2010ra}
\begin{equation}\label{eq:FFpara}
F^{XT}_{0}(q^{2}) = \frac{F^{hT}_{0}}{ 1-a_{T}\frac{q^{2}}{M_{h}^{2}}+b_{T}\left(\frac{q^{2}}{M_{h}^{2}}\right)^{2}},
\end{equation}
where the corresponding parameters are given in table~\ref{tab:FFinCLFQ}.

\newpage

\bibliographystyle{JHEP}
\bibliography{main.bib}

\providecommand{\href}[2]{#2}\begingroup\raggedright\begin{thebibliography}{10}

\bibitem{Altarelli:2013tya}
G.~Altarelli, \emph{{Collider Physics within the Standard Model: a Primer}},
  \href{https://arxiv.org/abs/1303.2842}{{\ttfamily 1303.2842}}.

\bibitem{Steigman:1976ev}
G.~Steigman, \emph{{Observational tests of antimatter cosmologies}},
  \href{https://doi.org/10.1146/annurev.aa.14.090176.002011}{\emph{Ann. Rev.
  Astron. Astrophys.} {\bfseries 14} (1976) 339}.

\bibitem{Riotto:1999yt}
A.~Riotto and M.~Trodden, \emph{{Recent progress in baryogenesis}},
  \href{https://doi.org/10.1146/annurev.nucl.49.1.35}{\emph{Ann. Rev. Nucl.
  Part. Sci.} {\bfseries 49} (1999) 35}
  [\href{https://arxiv.org/abs/hep-ph/9901362}{{\ttfamily hep-ph/9901362}}].

\bibitem{Canetti:2012zc}
L.~Canetti, M.~Drewes and M.~Shaposhnikov, \emph{{Matter and Antimatter in the
  Universe}}, \href{https://doi.org/10.1088/1367-2630/14/9/095012}{\emph{New J.
  Phys.} {\bfseries 14} (2012) 095012}
  [\href{https://arxiv.org/abs/1204.4186}{{\ttfamily 1204.4186}}].

\bibitem{Peebles:2013hla}
P.~J.~E. Peebles, \emph{{Dark Matter}},
  \href{https://doi.org/10.1073/pnas.1308786111}{\emph{Proc. Nat. Acad. Sci.}
  {\bfseries 112} (2015) 2246}
  [\href{https://arxiv.org/abs/1305.6859}{{\ttfamily 1305.6859}}].

\bibitem{Lukovic:2014vma}
V.~Lukovic, P.~Cabella and N.~Vittorio, \emph{{Dark matter in cosmology}},
  \href{https://doi.org/10.1142/S0217751X14430015}{\emph{Int. J. Mod. Phys. A}
  {\bfseries 29} (2014) 1443001}
  [\href{https://arxiv.org/abs/1411.3556}{{\ttfamily 1411.3556}}].

\bibitem{Bertone:2016nfn}
G.~Bertone and D.~Hooper, \emph{{History of dark matter}},
  \href{https://doi.org/10.1103/RevModPhys.90.045002}{\emph{Rev. Mod. Phys.}
  {\bfseries 90} (2018) 045002}
  [\href{https://arxiv.org/abs/1605.04909}{{\ttfamily 1605.04909}}].

\bibitem{Bilenky:1987ty}
S.~M. Bilenky and S.~T. Petcov, \emph{{Massive Neutrinos and Neutrino
  Oscillations}}, \href{https://doi.org/10.1103/RevModPhys.59.671}{\emph{Rev.
  Mod. Phys.} {\bfseries 59} (1987) 671}.

\bibitem{Strumia:2006db}
A.~Strumia and F.~Vissani, \emph{{Neutrino masses and mixings and...}},
  \href{https://arxiv.org/abs/hep-ph/0606054}{{\ttfamily hep-ph/0606054}}.

\bibitem{deSalas:2017kay}
P.~F. de~Salas, D.~V. Forero, C.~A. Ternes, M.~Tortola and J.~W.~F. Valle,
  \emph{{Status of neutrino oscillations 2018: 3$\sigma$ hint for normal mass
  ordering and improved CP sensitivity}},
  \href{https://doi.org/10.1016/j.physletb.2018.06.019}{\emph{Phys. Lett. B}
  {\bfseries 782} (2018) 633}
  [\href{https://arxiv.org/abs/1708.01186}{{\ttfamily 1708.01186}}].

\bibitem{Czarnecki:1997bu}
A.~Czarnecki and B.~Krause, \emph{{Neutron electric dipole moment in the
  standard model: Valence quark contributions}},
  \href{https://doi.org/10.1103/PhysRevLett.78.4339}{\emph{Phys. Rev. Lett.}
  {\bfseries 78} (1997) 4339}
  [\href{https://arxiv.org/abs/hep-ph/9704355}{{\ttfamily hep-ph/9704355}}].

\bibitem{Kim:2008hd}
J.~E. Kim and G.~Carosi, \emph{{Axions and the Strong CP Problem}},
  \href{https://doi.org/10.1103/RevModPhys.82.557}{\emph{Rev. Mod. Phys.}
  {\bfseries 82} (2010) 557} [\href{https://arxiv.org/abs/0807.3125}{{\ttfamily
  0807.3125}}].

\bibitem{Schmaltz:2005ky}
M.~Schmaltz and D.~Tucker-Smith, \emph{{Little Higgs review}},
  \href{https://doi.org/10.1146/annurev.nucl.55.090704.151502}{\emph{Ann. Rev.
  Nucl. Part. Sci.} {\bfseries 55} (2005) 229}
  [\href{https://arxiv.org/abs/hep-ph/0502182}{{\ttfamily hep-ph/0502182}}].

\bibitem{Wells:2016luz}
J.~D. Wells, \emph{{Higgs naturalness and the scalar boson proliferation
  instability problem}},
  \href{https://doi.org/10.1007/s11229-014-0618-8}{\emph{Synthese} {\bfseries
  194} (2017) 477} [\href{https://arxiv.org/abs/1603.06131}{{\ttfamily
  1603.06131}}].

\bibitem{Degrassi:2012ry}
G.~Degrassi, S.~Di~Vita, J.~Elias-Miro, J.~R. Espinosa, G.~F. Giudice,
  G.~Isidori et~al., \emph{{Higgs mass and vacuum stability in the Standard
  Model at NNLO}}, \href{https://doi.org/10.1007/JHEP08(2012)098}{\emph{JHEP}
  {\bfseries 08} (2012) 098} [\href{https://arxiv.org/abs/1205.6497}{{\ttfamily
  1205.6497}}].

\bibitem{Bezrukov:2014ina}
F.~Bezrukov and M.~Shaposhnikov, \emph{{Why should we care about the top quark
  Yukawa coupling?}}, \href{https://doi.org/10.1134/S1063776115030152}{\emph{J.
  Exp. Theor. Phys.} {\bfseries 120} (2015) 335}
  [\href{https://arxiv.org/abs/1411.1923}{{\ttfamily 1411.1923}}].

\bibitem{Padmanabhan:2002ji}
T.~Padmanabhan, \emph{{Cosmological constant: The Weight of the vacuum}},
  \href{https://doi.org/10.1016/S0370-1573(03)00120-0}{\emph{Phys. Rept.}
  {\bfseries 380} (2003) 235}
  [\href{https://arxiv.org/abs/hep-th/0212290}{{\ttfamily hep-th/0212290}}].

\bibitem{Rubakov:2017xzr}
V.~A. Rubakov and D.~S. Gorbunov, \emph{{Introduction to the Theory of the
  Early Universe}: {Hot big bang theory}}. World Scientific, Singapore, 2017,
  \href{https://doi.org/10.1142/10447}{10.1142/10447}.

\bibitem{Golling:2016gvc}
T.~Golling et~al., \emph{{Physics at a 100 TeV pp collider: beyond the Standard
  Model phenomena}},  \href{https://arxiv.org/abs/1606.00947}{{\ttfamily
  1606.00947}}.

\bibitem{FCC:2018byv}
{\scshape FCC} collaboration, A.~Abada et~al., \emph{{FCC Physics
  Opportunities}: {Future Circular Collider Conceptual Design Report Volume
  1}}, \href{https://doi.org/10.1140/epjc/s10052-019-6904-3}{\emph{Eur. Phys.
  J. C} {\bfseries 79} (2019) 474}.

\bibitem{Beacham:2019nyx}
J.~Beacham et~al., \emph{{Physics Beyond Colliders at CERN: Beyond the Standard
  Model Working Group Report}},
  \href{https://doi.org/10.1088/1361-6471/ab4cd2}{\emph{J. Phys. G} {\bfseries
  47} (2020) 010501} [\href{https://arxiv.org/abs/1901.09966}{{\ttfamily
  1901.09966}}].

\bibitem{Lanfranchi:2020crw}
G.~Lanfranchi, M.~Pospelov and P.~Schuster, \emph{{The Search for Feebly
  Interacting Particles}},
  \href{https://doi.org/10.1146/annurev-nucl-102419-055056}{\emph{Ann. Rev.
  Nucl. Part. Sci.} {\bfseries 71} (2021) 279}
  [\href{https://arxiv.org/abs/2011.02157}{{\ttfamily 2011.02157}}].

\bibitem{Curtin:2018mvb}
D.~Curtin et~al., \emph{{Long-Lived Particles at the Energy Frontier: The
  MATHUSLA Physics Case}},
  \href{https://doi.org/10.1088/1361-6633/ab28d6}{\emph{Rept. Prog. Phys.}
  {\bfseries 82} (2019) 116201}
  [\href{https://arxiv.org/abs/1806.07396}{{\ttfamily 1806.07396}}].

\bibitem{Cerci:2021nlb}
S.~Cerci et~al., \emph{{FACET: A new long-lived particle detector in the very
  forward region of the CMS experiment}},
  \href{https://arxiv.org/abs/2201.00019}{{\ttfamily 2201.00019}}.

\bibitem{FASER:2018ceo}
{\scshape FASER} collaboration, A.~Ariga et~al., \emph{{Letter of Intent for
  FASER: ForwArd Search ExpeRiment at the LHC}},
  \href{https://arxiv.org/abs/1811.10243}{{\ttfamily 1811.10243}}.

\bibitem{FASER:2018eoc}
{\scshape FASER} collaboration, A.~Ariga et~al., \emph{{FASER\textquoteright{}s
  physics reach for long-lived particles}},
  \href{https://doi.org/10.1103/PhysRevD.99.095011}{\emph{Phys. Rev. D}
  {\bfseries 99} (2019) 095011}
  [\href{https://arxiv.org/abs/1811.12522}{{\ttfamily 1811.12522}}].

\bibitem{Anelli:2015pba}
{\scshape SHiP} collaboration, M.~Anelli et~al., \emph{{A facility to Search
  for Hidden Particles (SHiP) at the CERN SPS}},
  \href{https://arxiv.org/abs/1504.04956}{{\ttfamily 1504.04956}}.

\bibitem{Alekhin:2015byh}
S.~Alekhin et~al., \emph{{A facility to Search for Hidden Particles at the CERN
  SPS: the SHiP physics case}},
  \href{https://doi.org/10.1088/0034-4885/79/12/124201}{\emph{Rept. Prog.
  Phys.} {\bfseries 79} (2016) 124201}
  [\href{https://arxiv.org/abs/1504.04855}{{\ttfamily 1504.04855}}].

\bibitem{Mermod:2017ceo}
{\scshape SHiP} collaboration, P.~Mermod, \emph{{Prospects of the SHiP and NA62
  experiments at CERN for hidden sector searches}},
  \href{https://doi.org/10.22323/1.295.0139}{\emph{PoS} {\bfseries NuFact2017}
  (2017) 139} [\href{https://arxiv.org/abs/1712.01768}{{\ttfamily
  1712.01768}}].

\bibitem{NA62:2017qcd}
{\scshape NA62} collaboration, E.~Cortina~Gil et~al., \emph{{Search for heavy
  neutral lepton production in $K^+$ decays}},
  \href{https://doi.org/10.1016/j.physletb.2018.01.031}{\emph{Phys. Lett. B}
  {\bfseries 778} (2018) 137}
  [\href{https://arxiv.org/abs/1712.00297}{{\ttfamily 1712.00297}}].

\bibitem{Drewes:2018gkc}
M.~Drewes, J.~Hajer, J.~Klaric and G.~Lanfranchi, \emph{{NA62 sensitivity to
  heavy neutral leptons in the low scale seesaw model}},
  \href{https://doi.org/10.1007/JHEP07(2018)105}{\emph{JHEP} {\bfseries 07}
  (2018) 105} [\href{https://arxiv.org/abs/1801.04207}{{\ttfamily
  1801.04207}}].

\bibitem{DUNE:2015lol}
{\scshape DUNE} collaboration, R.~Acciarri et~al., \emph{{Long-Baseline
  Neutrino Facility (LBNF) and Deep Underground Neutrino Experiment (DUNE)}:
  {Conceptual Design Report, Volume 2: The Physics Program for DUNE at LBNF}},
  \href{https://arxiv.org/abs/1512.06148}{{\ttfamily 1512.06148}}.

\bibitem{DUNE:2020fgq}
{\scshape DUNE} collaboration, B.~Abi et~al., \emph{{Prospects for beyond the
  Standard Model physics searches at the Deep Underground Neutrino
  Experiment}},
  \href{https://doi.org/10.1140/epjc/s10052-021-09007-w}{\emph{Eur. Phys. J. C}
  {\bfseries 81} (2021) 322}
  [\href{https://arxiv.org/abs/2008.12769}{{\ttfamily 2008.12769}}].

\bibitem{Patt:2006fw}
B.~Patt and F.~Wilczek, \emph{{Higgs-field portal into hidden sectors}},
  \href{https://arxiv.org/abs/hep-ph/0605188}{{\ttfamily hep-ph/0605188}}.

\bibitem{Bezrukov:2009yw}
F.~Bezrukov and D.~Gorbunov, \emph{{Light inflaton Hunter's Guide}},
  \href{https://doi.org/10.1007/JHEP05(2010)010}{\emph{JHEP} {\bfseries 05}
  (2010) 010} [\href{https://arxiv.org/abs/0912.0390}{{\ttfamily 0912.0390}}].

\bibitem{Boiarska:2019jym}
I.~Boiarska, K.~Bondarenko, A.~Boyarsky, V.~Gorkavenko, M.~Ovchynnikov and
  A.~Sokolenko, \emph{{Phenomenology of GeV-scale scalar portal}},
  \href{https://doi.org/10.1007/JHEP11(2019)162}{\emph{JHEP} {\bfseries 11}
  (2019) 162} [\href{https://arxiv.org/abs/1904.10447}{{\ttfamily
  1904.10447}}].

\bibitem{Peccei:1977hh}
R.~D. Peccei and H.~R. Quinn, \emph{{CP Conservation in the Presence of
  Instantons}}, \href{https://doi.org/10.1103/PhysRevLett.38.1440}{\emph{Phys.
  Rev. Lett.} {\bfseries 38} (1977) 1440}.

\bibitem{Weinberg:1977ma}
S.~Weinberg, \emph{{A New Light Boson?}},
  \href{https://doi.org/10.1103/PhysRevLett.40.223}{\emph{Phys. Rev. Lett.}
  {\bfseries 40} (1978) 223}.

\bibitem{Wilczek:1977pj}
F.~Wilczek, \emph{{Problem of Strong $P$ and $T$ Invariance in the Presence of
  Instantons}}, \href{https://doi.org/10.1103/PhysRevLett.40.279}{\emph{Phys.
  Rev. Lett.} {\bfseries 40} (1978) 279}.

\bibitem{Choi:2020rgn}
K.~Choi, S.~H. Im and C.~S. Shin, \emph{{Recent progress in physics of axions
  or axion-like particles}},
  \href{https://arxiv.org/abs/2012.05029}{{\ttfamily 2012.05029}}.

\bibitem{Okun:1982xi}
L.~B. Okun, \emph{{LIMITS OF ELECTRODYNAMICS: PARAPHOTONS?}}, {\emph{Sov. Phys.
  JETP} {\bfseries 56} (1982) 502}.

\bibitem{Holdom:1985ag}
B.~Holdom, \emph{{Two U(1)'s and Epsilon Charge Shifts}},
  \href{https://doi.org/10.1016/0370-2693(86)91377-8}{\emph{Phys. Lett. B}
  {\bfseries 166} (1986) 196}.

\bibitem{Langacker:2008yv}
P.~Langacker, \emph{{The Physics of Heavy $Z^\prime$ Gauge Bosons}},
  \href{https://doi.org/10.1103/RevModPhys.81.1199}{\emph{Rev. Mod. Phys.}
  {\bfseries 81} (2009) 1199}
  [\href{https://arxiv.org/abs/0801.1345}{{\ttfamily 0801.1345}}].

\bibitem{Asaka:2005pn}
T.~Asaka and M.~Shaposhnikov, \emph{{The $\nu$MSM, dark matter and baryon
  asymmetry of the universe}},
  \href{https://doi.org/10.1016/j.physletb.2005.06.020}{\emph{Phys. Lett. B}
  {\bfseries 620} (2005) 17}
  [\href{https://arxiv.org/abs/hep-ph/0505013}{{\ttfamily hep-ph/0505013}}].

\bibitem{Asaka:2005an}
T.~Asaka, S.~Blanchet and M.~Shaposhnikov, \emph{{The nuMSM, dark matter and
  neutrino masses}},
  \href{https://doi.org/10.1016/j.physletb.2005.09.070}{\emph{Phys. Lett. B}
  {\bfseries 631} (2005) 151}
  [\href{https://arxiv.org/abs/hep-ph/0503065}{{\ttfamily hep-ph/0503065}}].

\bibitem{Bondarenko:2018ptm}
K.~Bondarenko, A.~Boyarsky, D.~Gorbunov and O.~Ruchayskiy, \emph{{Phenomenology
  of GeV-scale Heavy Neutral Leptons}},
  \href{https://doi.org/10.1007/JHEP11(2018)032}{\emph{JHEP} {\bfseries 11}
  (2018) 032} [\href{https://arxiv.org/abs/1805.08567}{{\ttfamily
  1805.08567}}].

\bibitem{Boyarsky:2018tvu}
A.~Boyarsky, M.~Drewes, T.~Lasserre, S.~Mertens and O.~Ruchayskiy,
  \emph{{Sterile neutrino Dark Matter}},
  \href{https://doi.org/10.1016/j.ppnp.2018.07.004}{\emph{Prog. Part. Nucl.
  Phys.} {\bfseries 104} (2019) 1}
  [\href{https://arxiv.org/abs/1807.07938}{{\ttfamily 1807.07938}}].

\bibitem{Antoniadis:2009ze}
I.~Antoniadis, A.~Boyarsky, S.~Espahbodi, O.~Ruchayskiy and J.~D. Wells,
  \emph{{Anomaly driven signatures of new invisible physics at the Large Hadron
  Collider}},
  \href{https://doi.org/10.1016/j.nuclphysb.2009.09.009}{\emph{Nucl. Phys. B}
  {\bfseries 824} (2010) 296}
  [\href{https://arxiv.org/abs/0901.0639}{{\ttfamily 0901.0639}}].

\bibitem{DHoker:1984izu}
E.~D'Hoker and E.~Farhi, \emph{{Decoupling a Fermion Whose Mass Is Generated by
  a Yukawa Coupling: The General Case}},
  \href{https://doi.org/10.1016/0550-3213(84)90586-8}{\emph{Nucl. Phys. B}
  {\bfseries 248} (1984) 59}.

\bibitem{DHoker:1984mif}
E.~D'Hoker and E.~Farhi, \emph{{Decoupling a Fermion in the Standard
  Electroweak Theory}},
  \href{https://doi.org/10.1016/0550-3213(84)90587-X}{\emph{Nucl. Phys. B}
  {\bfseries 248} (1984) 77}.

\bibitem{Antoniadis:2000ena}
I.~Antoniadis, E.~Kiritsis and T.~N. Tomaras, \emph{{A D-brane alternative to
  unification}},
  \href{https://doi.org/10.1016/S0370-2693(00)00733-4}{\emph{Phys. Lett. B}
  {\bfseries 486} (2000) 186}
  [\href{https://arxiv.org/abs/hep-ph/0004214}{{\ttfamily hep-ph/0004214}}].

\bibitem{Coriano:2005own}
C.~Coriano, N.~Irges and E.~Kiritsis, \emph{{On the effective theory of low
  scale orientifold string vacua}},
  \href{https://doi.org/10.1016/j.nuclphysb.2006.04.009}{\emph{Nucl. Phys. B}
  {\bfseries 746} (2006) 77}
  [\href{https://arxiv.org/abs/hep-ph/0510332}{{\ttfamily hep-ph/0510332}}].

\bibitem{Coriano:2007xg}
N.~Irges, C.~Coriano and S.~Morelli, \emph{{Stuckelberg Axions and the
  Effective Action of Anomalous Abelian Models 2. A SU(3)C x SU(2)W x U(1)Y x
  U(1)B model and its signature at the LHC}},
  \href{https://doi.org/10.1016/j.nuclphysb.2007.07.027}{\emph{Nucl. Phys. B}
  {\bfseries 789} (2008) 133}
  [\href{https://arxiv.org/abs/hep-ph/0703127}{{\ttfamily hep-ph/0703127}}].

\bibitem{Anastasopoulos:2006cz}
P.~Anastasopoulos, M.~Bianchi, E.~Dudas and E.~Kiritsis, \emph{{Anomalies,
  anomalous U(1)'s and generalized Chern-Simons terms}},
  \href{https://doi.org/10.1088/1126-6708/2006/11/057}{\emph{JHEP} {\bfseries
  11} (2006) 057} [\href{https://arxiv.org/abs/hep-th/0605225}{{\ttfamily
  hep-th/0605225}}].

\bibitem{Anastasopoulos:2008jt}
P.~Anastasopoulos, F.~Fucito, A.~Lionetto, G.~Pradisi, A.~Racioppi and Y.~S.
  Stanev, \emph{{Minimal Anomalous U(1)-prime Extension of the MSSM}},
  \href{https://doi.org/10.1103/PhysRevD.78.085014}{\emph{Phys. Rev. D}
  {\bfseries 78} (2008) 085014}
  [\href{https://arxiv.org/abs/0804.1156}{{\ttfamily 0804.1156}}].

\bibitem{Harvey:2007ca}
J.~A. Harvey, C.~T. Hill and R.~J. Hill, \emph{{Standard Model Gauging of the
  Wess-Zumino-Witten Term: Anomalies, Global Currents and pseudo-Chern-Simons
  Interactions}}, \href{https://doi.org/10.1103/PhysRevD.77.085017}{\emph{Phys.
  Rev. D} {\bfseries 77} (2008) 085017}
  [\href{https://arxiv.org/abs/0712.1230}{{\ttfamily 0712.1230}}].

\bibitem{Dudas:2009uq}
E.~Dudas, Y.~Mambrini, S.~Pokorski and A.~Romagnoni, \emph{{(In)visible Z-prime
  and dark matter}},
  \href{https://doi.org/10.1088/1126-6708/2009/08/014}{\emph{JHEP} {\bfseries
  08} (2009) 014} [\href{https://arxiv.org/abs/0904.1745}{{\ttfamily
  0904.1745}}].

\bibitem{Kumar:2007zza}
J.~Kumar, A.~Rajaraman and J.~D. Wells, \emph{{Probing the Green-Schwarz
  Mechanism at the Large Hadron Collider}},
  \href{https://doi.org/10.1103/PhysRevD.77.066011}{\emph{Phys. Rev. D}
  {\bfseries 77} (2008) 066011}
  [\href{https://arxiv.org/abs/0707.3488}{{\ttfamily 0707.3488}}].

\bibitem{Dror:2017ehi}
J.~A. Dror, R.~Lasenby and M.~Pospelov, \emph{{New constraints on light vectors
  coupled to anomalous currents}},
  \href{https://doi.org/10.1103/PhysRevLett.119.141803}{\emph{Phys. Rev. Lett.}
  {\bfseries 119} (2017) 141803}
  [\href{https://arxiv.org/abs/1705.06726}{{\ttfamily 1705.06726}}].

\bibitem{Dror:2017nsg}
J.~A. Dror, R.~Lasenby and M.~Pospelov, \emph{{Dark forces coupled to
  nonconserved currents}},
  \href{https://doi.org/10.1103/PhysRevD.96.075036}{\emph{Phys. Rev. D}
  {\bfseries 96} (2017) 075036}
  [\href{https://arxiv.org/abs/1707.01503}{{\ttfamily 1707.01503}}].

\bibitem{Boyarsky:2005hs}
A.~Boyarsky, O.~Ruchayskiy and M.~Shaposhnikov, \emph{{Observational
  manifestations of anomaly inflow}},
  \href{https://doi.org/10.1103/PhysRevD.72.085011}{\emph{Phys. Rev. D}
  {\bfseries 72} (2005) 085011}
  [\href{https://arxiv.org/abs/hep-th/0507098}{{\ttfamily hep-th/0507098}}].

\bibitem{Boyarsky:2005eq}
A.~Boyarsky, O.~Ruchayskiy and M.~Shaposhnikov, \emph{{Anomalies as a signature
  of extra dimensions}},
  \href{https://doi.org/10.1016/j.physletb.2005.08.084}{\emph{Phys. Lett. B}
  {\bfseries 626} (2005) 184}
  [\href{https://arxiv.org/abs/hep-ph/0507195}{{\ttfamily hep-ph/0507195}}].

\bibitem{Antoniadis:2006wp}
I.~Antoniadis, A.~Boyarsky and O.~Ruchayskiy, \emph{{Axion alternatives}},
  \href{https://arxiv.org/abs/hep-ph/0606306}{{\ttfamily hep-ph/0606306}}.

\bibitem{Antoniadis:2007sp}
I.~Antoniadis, A.~Boyarsky and O.~Ruchayskiy, \emph{{Anomaly induced effects in
  a magnetic field}},
  \href{https://doi.org/10.1016/j.nuclphysb.2007.10.006}{\emph{Nucl. Phys. B}
  {\bfseries 793} (2008) 246}
  [\href{https://arxiv.org/abs/0708.3001}{{\ttfamily 0708.3001}}].

\bibitem{Williams:1971ms}
E.~R. Williams, J.~E. Faller and H.~A. Hill, \emph{{New experimental test of
  Coulomb's law: A Laboratory upper limit on the photon rest mass}},
  \href{https://doi.org/10.1103/PhysRevLett.26.721}{\emph{Phys. Rev. Lett.}
  {\bfseries 26} (1971) 721}.

\bibitem{Tu:2005ge}
L.-C. Tu, J.~Luo and G.~T. Gillies, \emph{{The mass of the photon}},
  \href{https://doi.org/10.1088/0034-4885/68/1/R02}{\emph{Rept. Prog. Phys.}
  {\bfseries 68} (2005) 77}.

\bibitem{Davidson:1991si}
S.~Davidson, B.~Campbell and D.~C. Bailey, \emph{{Limits on particles of small
  electric charge}},
  \href{https://doi.org/10.1103/PhysRevD.43.2314}{\emph{Phys. Rev. D}
  {\bfseries 43} (1991) 2314}.

\bibitem{Davidson:2000hf}
S.~Davidson, S.~Hannestad and G.~Raffelt, \emph{{Updated bounds on millicharged
  particles}}, \href{https://doi.org/10.1088/1126-6708/2000/05/003}{\emph{JHEP}
  {\bfseries 05} (2000) 003}
  [\href{https://arxiv.org/abs/hep-ph/0001179}{{\ttfamily hep-ph/0001179}}].

\bibitem{Marinelli:1983nd}
M.~Marinelli and G.~Morpurgo, \emph{{The Electric Neutrality of Matter: A
  Summary}}, \href{https://doi.org/10.1016/0370-2693(84)91752-0}{\emph{Phys.
  Lett. B} {\bfseries 137} (1984) 439}.

\bibitem{Zyla:2020zbs}
{\scshape Particle Data Group} collaboration, P.~A. Zyla et~al., \emph{{Review
  of Particle Physics}},
  \href{https://doi.org/10.1093/ptep/ptaa104}{\emph{PTEP} {\bfseries 2020}
  (2020) 083C01}.

\bibitem{Hewett:1988xc}
J.~L. Hewett and T.~G. Rizzo, \emph{{Low-Energy Phenomenology of Superstring
  Inspired E(6) Models}},
  \href{https://doi.org/10.1016/0370-1573(89)90071-9}{\emph{Phys. Rept.}
  {\bfseries 183} (1989) 193}.

\bibitem{Gunion:1989we}
J.~F. Gunion, H.~E. Haber, G.~L. Kane and S.~Dawson, \emph{{The Higgs Hunter's
  Guide}}. CRC Press, Boca Raton, 1990,
  \href{https://doi.org/10.1201/9780429496448}{10.1201/9780429496448}.

\bibitem{Leutwyler:1989xj}
H.~Leutwyler and M.~A. Shifman, \emph{{Light Higgs Particle in Decays of $K$
  and $\eta$ Mesons}},
  \href{https://doi.org/10.1016/0550-3213(90)90475-S}{\emph{Nucl. Phys.}
  {\bfseries B343} (1990) 369}.

\bibitem{Kallen:1964lxa}
G.~K\"all\'en, \emph{{Elementary particle physics}}. Addison-Wesley, Reading,
  MA, 1964.

\bibitem{Bogolyubov:1983gp}
N.~N. Bogolyubov and D.~V. Shirkov, \emph{{QUANTUM FIELDS}}. Benjamin Cummings,
  San Francisco, 1983.

\bibitem{Ball:2004ye}
P.~Ball and R.~Zwicky, \emph{{New results on $B \to \pi, K, \eta$ decay
  formfactors from light-cone sum rules}},
  \href{https://doi.org/10.1103/PhysRevD.71.014015}{\emph{Phys. Rev.}
  {\bfseries D71} (2005) 014015}
  [\href{https://arxiv.org/abs/hep-ph/0406232}{{\ttfamily hep-ph/0406232}}].

\bibitem{Yushchenko:2004zs}
O.~P. Yushchenko et~al., \emph{{High statistic measurement of the K-
  ---\ensuremath{>} pi0 e- nu decay form-factors}},
  \href{https://doi.org/10.1016/j.physletb.2004.03.069}{\emph{Phys. Lett. B}
  {\bfseries 589} (2004) 111}
  [\href{https://arxiv.org/abs/hep-ex/0404030}{{\ttfamily hep-ex/0404030}}].

\bibitem{OKA:2017lbd}
{\scshape OKA} collaboration, O.~P. Yushchenko et~al., \emph{{$K_{e3}$ decay
  studies in OKA experiment}},
  \href{https://doi.org/10.1134/S0021364018030037}{\emph{JETP Lett.} {\bfseries
  107} (2018) 139} [\href{https://arxiv.org/abs/1708.09587}{{\ttfamily
  1708.09587}}].

\bibitem{NA482:2018rgv}
{\scshape NA48/2} collaboration, J.~R. Batley et~al., \emph{{Measurement of the
  form factors of charged kaon semileptonic decays}},
  \href{https://doi.org/10.1007/JHEP10(2018)150}{\emph{JHEP} {\bfseries 10}
  (2018) 150} [\href{https://arxiv.org/abs/1808.09041}{{\ttfamily
  1808.09041}}].

\bibitem{NA48:2004jcz}
{\scshape NA48} collaboration, A.~Lai et~al., \emph{{Measurement of K0(e3)
  form-factors}},
  \href{https://doi.org/10.1016/j.physletb.2004.08.076}{\emph{Phys. Lett. B}
  {\bfseries 604} (2004) 1}
  [\href{https://arxiv.org/abs/hep-ex/0410065}{{\ttfamily hep-ex/0410065}}].

\bibitem{Alexopoulos:2004sy}
{\scshape KTeV} collaboration, T.~Alexopoulos et~al., \emph{{Measurements of
  semileptonic K(L) decay form-factors}},
  \href{https://doi.org/10.1103/PhysRevD.70.092007}{\emph{Phys. Rev.}
  {\bfseries D70} (2004) 092007}
  [\href{https://arxiv.org/abs/hep-ex/0406003}{{\ttfamily hep-ex/0406003}}].

\bibitem{KLOE:2006kms}
{\scshape KLOE} collaboration, F.~Ambrosino et~al., \emph{{Measurement of the
  form-factor slopes for the decay $K(L) \to \pi^\pm e^\mp \nu$ with the KLOE
  detector}}, \href{https://doi.org/10.1016/j.physletb.2006.03.036}{\emph{Phys.
  Lett. B} {\bfseries 636} (2006) 166}
  [\href{https://arxiv.org/abs/hep-ex/0601038}{{\ttfamily hep-ex/0601038}}].

\bibitem{KLOE:2006vvm}
{\scshape KLOE} collaboration, F.~Ambrosino et~al., \emph{{Study of the
  branching ratio and charge asymmetry for the decay $K(s) \to \pi e \nu$ with
  the KLOE detector}},
  \href{https://doi.org/10.1016/j.physletb.2006.03.047}{\emph{Phys. Lett. B}
  {\bfseries 636} (2006) 173}
  [\href{https://arxiv.org/abs/hep-ex/0601026}{{\ttfamily hep-ex/0601026}}].

\bibitem{Sun:2010nv}
Y.-J. Sun, Z.-H. Li and T.~Huang, \emph{{$B_{(s)}\to S$ transitions in the
  light cone sum rules with the chiral current}},
  \href{https://doi.org/10.1103/PhysRevD.83.025024}{\emph{Phys. Rev.}
  {\bfseries D83} (2011) 025024}
  [\href{https://arxiv.org/abs/1011.3901}{{\ttfamily 1011.3901}}].

\bibitem{Daldrop:2012sr}
{\scshape ETM} collaboration, J.~O. Daldrop, C.~Alexandrou, M.~Dalla~Brida,
  M.~Gravina, L.~Scorzato, C.~Urbach et~al., \emph{{Lattice investigation of
  the tetraquark candidates a0(980) and kappa}},
  \href{https://doi.org/10.22323/1.164.0161}{\emph{PoS} {\bfseries LATTICE2012}
  (2012) 161} [\href{https://arxiv.org/abs/1211.5002}{{\ttfamily 1211.5002}}].

\bibitem{Cheng:2013fba}
H.-Y. Cheng, C.-K. Chua, K.-C. Yang and Z.-Q. Zhang, \emph{{Revisiting
  charmless hadronic B decays to scalar mesons}},
  \href{https://doi.org/10.1103/PhysRevD.87.114001}{\emph{Phys. Rev.}
  {\bfseries D87} (2013) 114001}
  [\href{https://arxiv.org/abs/1303.4403}{{\ttfamily 1303.4403}}].

\bibitem{Issadykov:2015iba}
A.~Issadykov, M.~A. Ivanov and S.~K. Sakhiyev, \emph{{Form factors of the
  B-S-transitions in the covariant quark model}},
  \href{https://doi.org/10.1103/PhysRevD.91.074007}{\emph{Phys. Rev.}
  {\bfseries D91} (2015) 074007}
  [\href{https://arxiv.org/abs/1502.05280}{{\ttfamily 1502.05280}}].

\bibitem{Ball:2004rg}
P.~Ball and R.~Zwicky, \emph{{$B_{d,s} \to \rho, \omega, K^*, \phi$ decay
  form-factors from light-cone sum rules revisited}},
  \href{https://doi.org/10.1103/PhysRevD.71.014029}{\emph{Phys. Rev.}
  {\bfseries D71} (2005) 014029}
  [\href{https://arxiv.org/abs/hep-ph/0412079}{{\ttfamily hep-ph/0412079}}].

\bibitem{Ebert:1997mg}
D.~Ebert, R.~N. Faustov and V.~O. Galkin, \emph{{Exclusive nonleptonic decays
  of B mesons}}, \href{https://doi.org/10.1103/PhysRevD.56.312}{\emph{Phys.
  Rev.} {\bfseries D56} (1997) 312}
  [\href{https://arxiv.org/abs/hep-ph/9701218}{{\ttfamily hep-ph/9701218}}].

\bibitem{Hatanaka:2009gb}
H.~Hatanaka and K.-C. Yang, \emph{{Radiative and Semileptonic B Decays
  Involving the Tensor Meson $K_{2}^{*}(1430)$ in the Standard Model and
  Beyond}}, \href{https://doi.org/10.1103/PhysRevD.79.114008}{\emph{Phys. Rev.}
  {\bfseries D79} (2009) 114008}
  [\href{https://arxiv.org/abs/0903.1917}{{\ttfamily 0903.1917}}].

\bibitem{Lu:2011jm}
C.-D. Lu and W.~Wang, \emph{{Analysis of $B\to K^*_J (\to K \pi) \mu^+\mu^-$ in
  the higher kaon resonance region}},
  \href{https://doi.org/10.1103/PhysRevD.85.034014}{\emph{Phys. Rev.}
  {\bfseries D85} (2012) 034014}
  [\href{https://arxiv.org/abs/1111.1513}{{\ttfamily 1111.1513}}].

\bibitem{Bashiry:2009wq}
V.~Bashiry, \emph{{Lepton polarization in $B \to K_1 \ell^+ \ell^-$ Decays}},
  \href{https://doi.org/10.1088/1126-6708/2009/06/062}{\emph{JHEP} {\bfseries
  06} (2009) 062} [\href{https://arxiv.org/abs/0902.2578}{{\ttfamily
  0902.2578}}].

\bibitem{Hatanaka:2008gu}
H.~Hatanaka and K.-C. Yang, \emph{{$K_{1}(1270)$-$K_{1}(1400)$ Mixing Angle and
  New-Physics Effects in $B\to K_{1} l^{+} l^{-}$ Decays}},
  \href{https://doi.org/10.1103/PhysRevD.78.074007}{\emph{Phys. Rev.}
  {\bfseries D78} (2008) 074007}
  [\href{https://arxiv.org/abs/0808.3731}{{\ttfamily 0808.3731}}].

\bibitem{Cheng:2010yd}
H.-Y. Cheng and K.-C. Yang, \emph{{Charmless Hadronic B Decays into a Tensor
  Meson}}, \href{https://doi.org/10.1103/PhysRevD.83.034001}{\emph{Phys. Rev.}
  {\bfseries D83} (2011) 034001}
  [\href{https://arxiv.org/abs/1010.3309}{{\ttfamily 1010.3309}}].

\bibitem{Li:2010ra}
R.-H. Li, C.-D. Lu and W.~Wang, \emph{{Branching ratios, forward-backward
  asymmetries and angular distributions of $B\to K_2^*l^+l^-$ in the standard
  model and new physics scenarios}},
  \href{https://doi.org/10.1103/PhysRevD.83.034034}{\emph{Phys. Rev.}
  {\bfseries D83} (2011) 034034}
  [\href{https://arxiv.org/abs/1012.2129}{{\ttfamily 1012.2129}}].

\end{thebibliography}\endgroup

\end{document}